\documentclass[useAMS,usenatbib]{mn2e}

\usepackage{graphicx}
\usepackage{natbib}
\usepackage{eufrak}
\usepackage{color,graphicx}

\newcommand{\ud}{\rmn{d}}

\title[A Galactic Ring of Minimum Stellar Density]{A Galactic Ring of Minimum Stellar Density Near the Solar Orbit Radius}
\author[D. A. Barros et al.]{D. A. Barros\thanks{E-mail: 
douglasab@astro.iag.usp.br}, J. R. D. L\'epine\thanks{E-mail: jacques@astro.iag.usp.br}, T. C. Junqueira
\\
Instituto de Astronomia, Geof\'isica e Ci\^encias Atmosf\'ericas, Universidade de S\~ao Paulo, Cidade Universit\'aria, \\
S\~ao Paulo 05508-090, SP, Brasil}

\begin{document}

\date{Accepted. Received; in original form}

\pagerange{\pageref{firstpage}--\pageref{lastpage}} \pubyear{2013}

\maketitle

\label{firstpage}

\begin{abstract}

We analyse the secular effects of a long-lived Galactic spiral structure on the stellar orbits with mean radii close to the corotation resonance. By test-particle simulations and different spiral potential models with parameters constrained on observations, we verified the formation of a minimum with amplitude $\sim 30\%-40\%$ of the background disk stellar density at corotation.  
Such minimum is formed by the secular angular momentum transfer between stars and the spiral density wave on both sides of corotation. 
We demonstrate that the secular loss (gain) of angular momentum and decrease (increase) of mean orbital radius of stars just inside (outside) corotation can counterbalance the opposite trend of exchange of angular momentum shown by stars orbiting the librational points $L_{4/5}$ at the corotation circle. Such secular processes actually allow steady spiral waves to promote radial migration across corotation.
We propose some observational evidences for the minimum stellar density in the Galactic disk, such as its direct relation to the minimum in the observed rotation curve of the Galaxy at the radius $r\sim 9$ kpc (for $R_{0}=7.5$ kpc), as well as its association with a minimum in the distribution of Galactic radii of a sample of open clusters older than 1 Gyr. 
The closeness of the solar orbit radius to the corotation resonance implies that the solar orbit lies inside a ring of minimum surface density (stellar + gas).  
This also implies in a correction to larger values for the estimated total mass of the Galactic disk, and consequently, a greater contribution of the disk component to the inner rotation curve of the Galaxy.

\end{abstract}




\section{Introduction}
\label{intro}

The effect of resonances between the rotation period of stars of a galactic disc around the center of the galaxy and the epicycle period (the oscillations of a star with respect to its unperturbed circular orbit), has not yet been fully explored. The classical theory of spiral arms of \citet{Lin_Shu1964}, as well as other theories which consider that the spiral arms are caused by the crowding of stellar orbits (e.g. \citealt{Kalnajs1973,Pichardo2003,Junqueira2013}), consider that the spiral arm pattern rotates like a rigid body, with its own angular velocity noted $\Omega_p$. In this hypothesis, since the rotation curves of galaxies are relatively flat, and the linear velocity of a rigid body increases linearly with the radius ($v=\Omega_{p}\times r$), there is a point where the two lines intersect, and the material of the disk rotates at the same velocity of the pattern. This corotation point is of special interest in the theory of galactic structure, as will be seen in the present paper. It is also a resonance ($\Omega=\Omega_{p}$), although not of the same nature of the other ones, since it does not involve the epicycle frequency.
  
It must be said, however, that the classical theories of spiral arms to which we refer have not been able to predict the value of $\Omega_p$, which is treated as a free parameter to be determined by observations. If one adopts a value of $\Omega_p$ for a galaxy with a known rotation curve, then the position of all the resonances are automatically determined. Since $\Omega_p$ is not known a priori, the position of the resonances may seem to be arbitrary. In addition to this challenge, there are theories of spiral arms claiming that the pattern speed is variable, or that there are several pattern speeds superimposed (e.g. \citealt*{Sellwood_Binney2002,Merrifield2006}, among others), or even that the spiral arms are stochastic phenomena. The situation has changed more recently, when several measurable effects of corotation were observed in our Galaxy. There were several papers in the past which showed that the corotation radius is close to the solar orbit (e.g. \citealt*{Marochnik1972,Creze_Mennessier1973,Mishurov_Zenina1999}, among others). A direct measurement was made by \citet{Dias_Lepine2005} by using a sample of open clusters. These objects have known distances, ages and space velocities (proper motion and radial velocities), so that one can integrate their orbits towards the past and find their birthplaces. Since the clusters are born in spiral arms, one can follow the position of the spiral arms as a function of time. Based on this idea, \citet{Dias_Lepine2005} determined  the corotation radius equal to $(1.06\pm 0.08)R_{0}$, where $R_{0}$ is the distance of the Sun to the Galactic center. Throughout this paper, the subscript `$_{0}$' denotes the values of the parameters at the solar orbit radius.
 
A prediction of hydrodynamic analytical solutions and of hydrodynamic simulations is that a ring void of gas should form at the corotation radius, and indeed, the existence of this gap was evidenced by \citet*[hereafter ALM]{Amores2009} using the H\,{\sevensize I} LAB survey database (\citealt{Kalberla2005}). As discussed by ALM, the  Cassini-type gap had already been observed previously, but since it was not understood, no  attention was paid to it. An expected consequence of this ring void of gas is that there should be a depletion of young stars at the same radius, since the star-formation rate is believed to depend on the density of interstellar gas. And indeed, a gap in the distribution of young open clusters, and also of Cepheids, is observed (ALM). More than that, there is also a step in the metallicity distribution of the open clusters, the Fe abundance being 0.3 dex lower beyond corotation than just before it (\citealt[hereafter L+8]{Lepine2011b}). The step is interpreted as due to the independent chemical evolution on both sides of the barrier produced by the Cassini-like gap in the gas distribution. This is an indication that the spiral structure is long-lived, since the corotation radius has to be constant for a few billion years to build up such a metallicity step. Up to the moment, we were considering that the ring with a void of gas was also void of young stars only, although it was shown by \citet*{Lepine2003} that stars of any age initially situated at corotation will not stay there, but will have a `wandering' radial  motion with amplitude of several kiloparsecs.

In the present work, we present a deeper investigation of the effects of the forces acting on stars near the corotation radius, and we predict the existence of a minimum of stellar density at this radius, based both on theoretical considerations and on numerical simulations, using different models for the gravitational potential perturbation due to the spiral structure. The formation of the minimum stellar density at corotation has its basis on the exchange of energy and angular momentum between the disk stars and the spiral wave, according to the theories developed by \citet{Lynden-Bell_Kalnajs1972} and \citet{Zhang1996,Zhang1998,Zhang1999}. The secular redistribution of the surface density, induced by the angular momentum transfer in disks of spiral galaxies, is achieved by the decrease of the mean orbital radius for stars inside corotation, and the increase of this quantity for stars outside corotation. From a simple point of view, the corotation resonance acts like a repeller on the stellar orbits. Going further, we re-examine the Galactic distribution of open clusters, focusing now on the older objects (ages greater than 1 Gyr), and we show that the minimum of stellar density near the corotation circle indeed exists. An attempt to see the presence of such minimum in the distribution of red clump stars has also been made, but more work is needed in this field of investigation. We also show the connection between this small-density ring with a local minimum in the rotation curve of the Milky Way just beyond $R_{0}$.
 
The organization of this paper is as follows: in $\S$~\ref{gal_model} we present the Galactic model based on analytical expressions for both the axisymmetric and the spiral potentials, and the resulting location in the Galactic disk of the main resonances. In $\S$~\ref{effec_pot} we give the theoretical background for the formation of the density minimum at corotation, and in $\S$~\ref{num_exp} we give the details of the test-particle simulations. In $\S$~\ref{result_analys} we analyse the results of the simulations and compare them with the predictions from the theory. In $\S$~\ref{observ_evid} we present some observational evidences that give support to the existence of the minimum of stellar density. 
Some addititonal remarks about the implications of the minimum density to the properties of the Galaxy, and in particular, the hypothesis of the solar orbit situated inside a ring of minimum density, are discussed in $\S$~\ref{add_remarks}. Concluding remarks can be found in $\S$~\ref{outro}.



\section[]{The Galactic potential}
\label{gal_model}

We present in this section an analytical description of the Galactic gravitational potential which is needed to investigate the stellar orbits in the disk. Expressions for both the axisymmetric component and the non-axisymmetric one (associated with the spiral arms), and correspondig numerical parameters are given. The Galactic potential will be the basis for both the theoretical discussion and of numerical experiments in the following sections.

\subsection{The axisymmetric potential}
\label{potaxis}

\begin{figure}
 \includegraphics[scale=0.47]{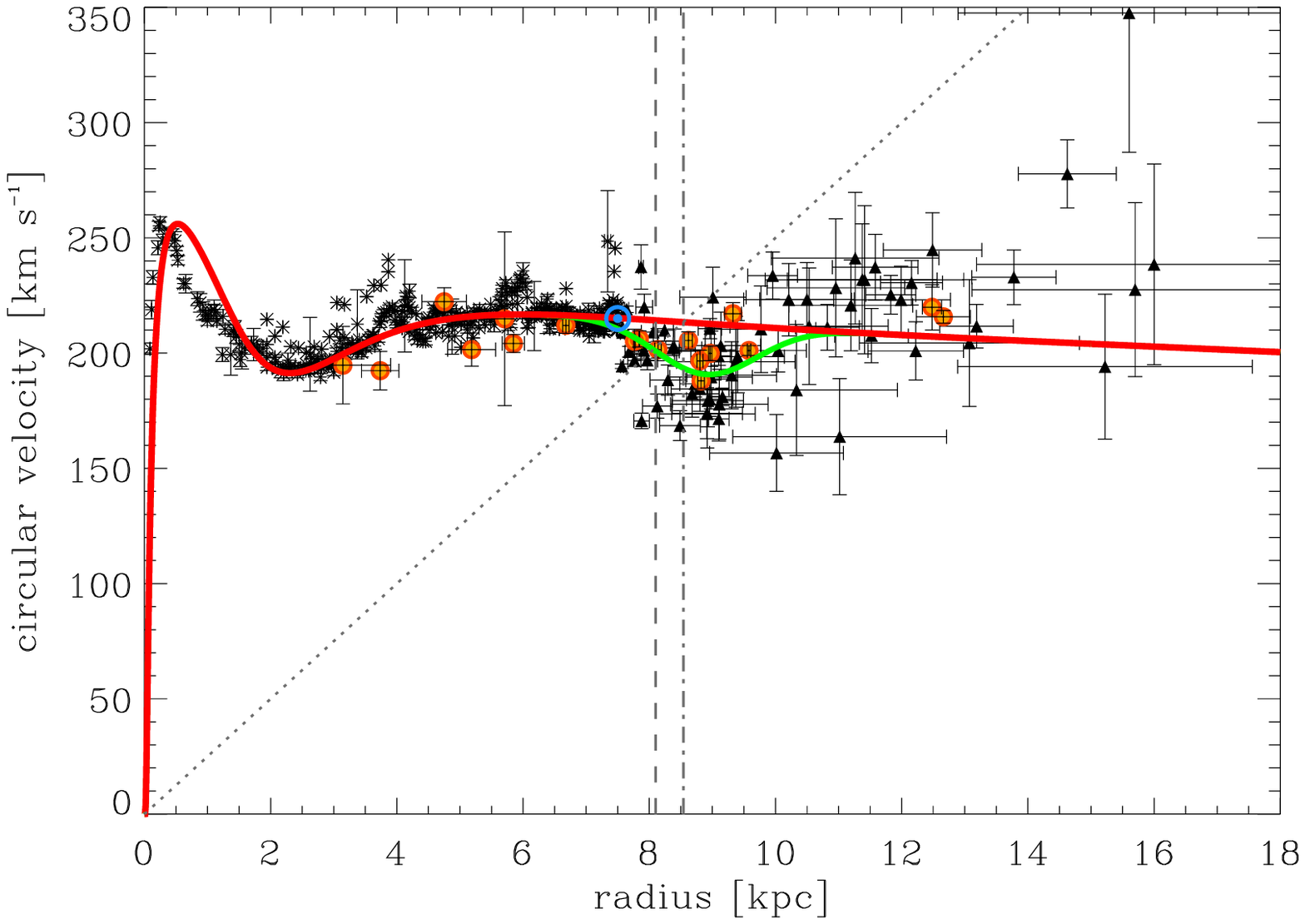}
 \caption{Observed rotation curve of the Galaxy. Asterisks: H\,{\sevensize I} and CO tangent velocities from \citet{Burton_Gordon1978} and \citet{Clemens1985} (compiled by \citealt{Sofue2009}); triangles: CO and H\,{\sevensize II} regions from \citet{Fich_Blitz_Stark1989} and \citet{Blitz_Fich_Stark1982}; filled orange circles: masers from high-mass star-forming regions from \citet{Reid2009}. The solid red curve is the plot of the function shown in Eq.~\ref{eq:v_L2011}; the green curve shows a Gaussian function (Eq.~\ref{eq:funct_mrc}) added to Eq.~\ref{eq:v_L2011} to fit the data located below the red curve. The dotted line indicates the linear velocity of the pattern $v=\Omega_{p}r$. The vertical dash-dotted and dashed lines indicate the corotation radius for the rotation curve without and with the Gaussian minimum added, respectively. The blue circle indicates the position of the Sun}
	\label{fig:rotcurve}
	\end{figure}

In order to be able to relate our findings to the structure of the Milky Way, we adopt a model for the axisymmetric galactic potential that reproduces the general behaviour of the rotation curve of the Galaxy. We use an analytical expression to represent the circular velocity as a function of Galactic radius, conveniently fitted by exponentials in the form (units are km s$^{-1}$ and kpc):

\begin{equation}
\label{eq:v_L2011}
V_{c}(r)=\alpha\exp\left[-\frac{r}{\beta}-\left(\frac{\gamma}{r}\right)^{2}\right]+\delta\exp\left[-\frac{r}{\epsilon}-\frac{\eta}{r}\right].
\end{equation}
The plot of the above expression is shown by the solid red curve in Fig.~\ref{fig:rotcurve}. As can be seen, the expression provides a quite smooth rotation curve, that gives the axisymmetric potential which will be used in the numerical experiments. For some examples throughout the paper, we also use a rotation curve formed by the function in Eq.~\ref{eq:v_L2011} to which a Gaussian minimum is added to fit the data in the observed rotation curve at radii $\sim 8-9.5$ kpc. The Gaussian function for the minimum in the rotation curve (mrc) is chosen as: 

\begin{equation}
f_{\mathrm{mrc}}(r)=-A_{\mathrm{mrc}}\exp\left[-\frac{1}{2}\left(\frac{r-R_{\mathrm{mrc}}}{\sigma_{\mathrm{mrc}}}\right)^{2}\right],
\label{eq:funct_mrc}
\end{equation}
where $A_{\mathrm{mrc}}$ is the amplitude and $\sigma_{\mathrm{mrc}}$ is the half-width of the minimum centered at the radius $R_{\mathrm{mrc}}$. The rotation curve formed by the sum of the expressions in Eqs.~\ref{eq:v_L2011} and~\ref{eq:funct_mrc}, $V_{c}(r)+f_{\mathrm{mrc}}(r)$, is close to that derived by \citet{Fich_Blitz_Stark1989} and is also similar to the ones previously used by our group (e.g. \citealt{Lepine2008}; ALM; \citealt{Lepine2011a}). The interpretation of a similar curve in terms of components of the Galaxy is given by \citet{Lepine_Leroy2000}. Since our interest here is only to adopt the best empirical rotation curve, we avoid any theoretical model or discussion on the mass components of the Galaxy. The green curve in Fig.~\ref{fig:rotcurve} is the plot of the Gaussian function in Eq.~\ref{eq:funct_mrc}. Table~\ref{tab:param_rotcurv} gives the values of the parameters chosen to reproduce the rotation curve of the Milky Way. 
For the Galactocentric distance of the Sun, we adopt $R_{0}$ = 7.5 kpc (for a detailed discussion about this choice, see \citealt{Lepine2008} and \citealt{Lepine2011a}). The circular velocity at $R_{0}$ resultant from Eq.~\ref{eq:v_L2011} is $V_{0}=215$ km s$^{-1}$. This value is chosen to satisfy the relation $V_{0}=R_{0}\Omega_{\odot}-v_{\odot}$, with the angular rotation velocity of the Sun $\Omega_{\odot}=30.24$ km s$^{-1}$ kpc$^{-1}$ (\citealt{Reid_Brunthaler2004}, based on the measurement of Sgr A$^{*}$ motion along the Galactic plane; also \citealt{Backer_Sramek1999}), and the peculiar velocity of the Sun in the direction of Galactic rotation $v_{\odot}=12.24$ km s$^{-1}$ (\citealt*{Schonrich2010}). Figure~\ref{fig:rotcurve} also shows the observed rotation curve of the Galaxy using several data obtained from the literature: H\,{\sevensize I} and CO tangent velocities from \citet{Burton_Gordon1978} and \citet{Clemens1985} (compiled by \citealt{Sofue2009}), for the inner Galaxy; CO and H\,{\sevensize II} regions from \citet{Fich_Blitz_Stark1989} and \citet{Blitz_Fich_Stark1982}, for the outer Galaxy; and maser emission from high-mass star-forming regions, from \citet{Reid2009}. All the rotation velocities and Galactocentric distances were re-calculated using the Galactic constants ($R_{0}$, $V_{0}$) adopted in this paper.  
Instead of being constant at large radii, the outer rotation curve of the Milky Way seems to present a slight falling of the circular velocity up to radius $\sim$ 60 kpc (\citealt{Xue2008}; \citealt{Sofue2012}). The rotation curve that we use in this paper also presents this general trend. However, as the main concern of this paper is about the structure of the stellar disk at the corotation radius, which is close to the solar orbit radius, the exact behaviour of the rotation curve of the Galaxy at large radii (and also in the central region, for radii smaller than $\sim 2$ kpc.)
must not affect our main results about the corotation resonance region.

As we restrict our study to orbits in the galactic plane, the axisymmetric potential can be derived directly from the rotation curve:

\begin{equation}
\label{eq:Phiax}
\Phi_{ax}(r)=\int{\frac{V^{2}_{c}}{r}\,\ud r}.
\end{equation}

\begin{table*}
 \centering
  \caption{Parameters of the rotation curve (Eqs.~\ref{eq:v_L2011} and~\ref{eq:funct_mrc}).}
  \label{tab:param_rotcurv}
  \begin{tabular}{@{}ccccccccc@{}}
  \hline
  \hline
   
  \multicolumn{1}{c}{$\balpha$} &
	\multicolumn{1}{c}{$\bbeta$} &
	\multicolumn{1}{c}{$\bgamma$} &
	\multicolumn{1}{c}{$\bdelta$} &
	\multicolumn{1}{c}{$\bepsilon$} &
	\multicolumn{1}{c}{$\boldeta$} &
	\multicolumn{1}{c}{$\mathbf{A_{\mathrm{mrc}}}$} & 
	\multicolumn{1}{c}{$\mathbf{R_{\mathrm{mrc}}}$} &
	\multicolumn{1}{c}{$\bsigma_{\mathrm{mrc}}$} \\
	
	\multicolumn{1}{c}{[km s$^{-1}$]} &
	\multicolumn{1}{c}{[kpc]} &
	\multicolumn{1}{c}{[kpc]} &
	\multicolumn{1}{c}{[km s$^{-1}$]} &
	\multicolumn{1}{c}{[kpc]} &
	\multicolumn{1}{c}{[kpc]} &
	\multicolumn{1}{c}{[km s$^{-1}$]} &
	\multicolumn{1}{c}{[kpc]} &
	\multicolumn{1}{c}{[kpc]} \\
 \hline
 240 &120 &3.4 &360 &3.1 &0.09 &22 &8.9 &0.8 \\
\hline
\end{tabular}
\end{table*} 

\subsection{The spiral potential}
\label{pot_non-axis}


For the gravitational potential due to the spiral perturbation, we use two different descriptions; one is the commonly employed analytical form:

\begin{equation}
\label{eq:potspiral_real}
\Phi_{sp}(r,\theta';t)=A_{sp}(r)\cos\left[\frac{m}{\tan i}\ln\left(\frac{r}{R_{0}}\right)-m(\theta'-\Omega_{p}t)\right]
\end{equation}
where $A_{sp}(r)$ is the amplitude of perturbation as a function of radius; {\it m\/} is the azimuthal wavenumber, which corresponds to the number of arms; {\it i\/} is the pitch angle of the spirals; and $\Omega_p$ is the pattern speed of the spiral structure. The potential is given in terms of the polar coordinates $r$ and $\theta'$ of the inertial reference frame, in which the spiral structure moves at the angular velocity $\Omega_{p}$.

The other description of the spiral potential is the new model recently proposed by \citet[hereafter JLBB]{Junqueira2013}. The potential is modeled as grooves with Gaussian profiles in directions normal to the arms, which are added to the axisymmetric potential of the disk. One of the advantages of this new potential is that its profile, instead of the cosine profile in the azimuthal direction of Eq.~\ref{eq:potspiral_real}, is more realistic when compared to observations. It is also more self-consistent, since the shape of the arms correspond to the density maxima formed by the crowding of periodic stellar orbits.
One of the main characteristics of this model is that it is composed only by potential wells, which are the grooves, whose effect is to add density in the arm region. This new potential is given by JLBB as:

\begin{equation}
\label{eq:potspiral_gauss}
\Phi_{sp_{g}}(r,\theta';t)=A_{sp_{g}}(r)\,e^{-\frac{r^2}{\sigma^2_g}\{1-\cos[\varphi(r)-m(\theta'-\Omega_{p}t)]\}},
\end{equation}
where $\sigma_{g}$ is the arm half-width in a direction tangent to the circle that crosses the arm at a given radius; 
$A_{sp_{g}}(r)$ is the amplitude of perturbation; and $\varphi(r)=\frac{m}{\tan i}\ln \left(\frac{r}{R_{0}}\right)$. The other parameters are equivalent to those given in Eq.~\ref{eq:potspiral_real}. 
    
Despite many observational efforts in the last decades, some properties of the Milky Way spiral structure remain not well determined. Regarding the geometry of the spiral arms, a number of authors have found different values for the total number of arms and pitch angle based on different tracers (\citealt{Vallee2002,Vallee2008}; \citealt{Melnik_Rautiainen2011}). \citet{Georgelin_Georgelin1976} proposed a 4-armed spiral pattern model with pitch angle $i\approx 12^{\circ}$ based on the distribution of H\,{\sevensize II} regions. This model has been updated by several authors (e.g. \citealt{Russeil2003}; \citealt{Paladini2004}; \citealt{Russeil2007}; \citealt{Hou2009}; \citealt{Efremov2011}). A structure with 4 spiral arms and pitch angle $i\approx 14^{\circ}$ was proposed by \citet{Ortiz_Lepine1993} in their model of the Galaxy for predicting star counts in the infrared. \citet{Drimmel_Spergel2001} identified 
a 2-armed stellar structure with pitch angle $i=17^{\circ}$ from the {\it COBE/DIRBE\/} {\it K\/} band emission profile of the Galactic plane; the 240 $\mu$m emission from dust in the interstellar gas is consistent with a 4-armed structure with pitch angle $i=13^{\circ}$. Two main stellar arms were identified by \citet{Churchwell2009} from the {\it Spitzer\/}/GLIMPSE infrared survey. Models composed by superpositions of two and four arms were previously used by \citet{Amaral_Lepine1997} and \citet{Lepine2001} to the study of self-consistency and structural parameters of the spiral pattern. Furthermore, it is known that some external galaxies do not show pure logarithmic spiral arms over long radial extensions (e.g. \citealt{Lepine2011a}, \citealt{Chernin1999}); some straight arms segments are observed in many cases, and this may be true for the Milky Way itself. There is still some disagreement between models that represent the arms in the outer regions of the Milky Way as an extrapolation of the observed inner arms (\citealt{Antoja2011}). Here, we adopt a geometric model, which we call `{\it basic model\/}', composed by four logarithmic spiral arms with pitch angle $i=-12^{\circ}$ (the negative sign corresponds to trailing spirals). Table~\ref{tab:models_non-axis} summarizes the characteristics of the models for the spiral structure that we used in the simulations, with the corresponding values of the parameters. The remaining parameters included in the table are next discussed. 
The last model (Sp6) in Table~\ref{tab:models_non-axis} includes the perturbation from a central bar in addition to that from the spiral structure. The parameters of the gravitational potential due to the central bar are described in $\S$~\ref{other_spiral_models} in conjunction with the results from the simulations using the model Sp6. 

\begin{table}
  \caption{Models of the non-axisymmetric component of the Galactic potential used in the simulations.}
  \label{tab:models_non-axis}
  \begin{tabular}{@{}ccl@{}}
  \hline
  \hline\\
     
  \multicolumn{1}{c}{\bf{Model}} &
	\multicolumn{1}{c}{\bf{Perturbation}} &
	\multicolumn{1}{l}{\bf{Characteristics}} \\
 \hline
	
 Sp1 &spirals &$m=4$ \\
     &cosine profile  &$i=-12^{\circ}$ \\
     &        &$f_{r0}=0.075$ \\
  \hline
  
 Sp2 &spirals &$m=2$ \\
     &cosine profile  &$i=-6^{\circ}$ \\
     &        &$f_{r0}=0.1$ \\
  \hline
  
 Sp3 &spirals with &$m=4$ \\
     &Gaussian profile &$i=-12^{\circ}$ \\
     &        &$\sigma_{g}=4.7$ kpc \\         
     &        &$\mathcal{A}_{\mathrm{sp_{g}}}=600$ \\
  \hline
  
 Sp4 &spirals &$m=2$ and $m=4$ \\
     &(two modes) &$i(m=2)=-6^{\circ}$ \\
     & cosine profile   &$i(m=4)=-12^{\circ}$ \\
     &        &$f_{r0}=0.05/0.075$ for both modes \\
     &        &$\Omega_{p}=25$ for both modes \\
  \hline
  
 Sp5 &spirals &$m=2$ and $m=4$ \\
     &(two modes) &$i=-12^{\circ}$ for both modes\\
     &cosine profile  &$f_{r0}=0.05/0.075$ for both modes \\
     &        &$\Omega_{p}(m=2)=18$ \\
     &        &$\Omega_{p}(m=4)=25$ \\
  \hline
  
 Sp6 &spirals          &spirals of model Sp1 \\
     &(cosine profile) &bar: $A_{b}=1250/2250$ \\
     &and central bar  &\ \ \ \ \ \ $\Omega_{b}=40$ \\
     &                 &\ \ \ \ \ \ $r_{b}=3.5$ kpc \\
  \hline

	\end{tabular}
	
	\medskip
	\scriptsize{
	The units are: $[\mathcal{A}_{\mathrm{sp_{g}}}]$ = km$^{2}$ s$^{-2}$ kpc$^{-1}$; $[A_{b}]$ = km$^{2}$ s$^{-2}$; $[\Omega_{p}]$ and $[\Omega_{b}]$ = km s$^{-1}$ kpc$^{-1}$. For all models, the scalelength of the arms is set as $r_{sp}=2.5$ kpc.}
	\end{table} 

Some different values for the pattern speed of the spiral structure of the Galaxy have been reported in the literature (see the review by \citealt{Gerhard2011}). In this paper, we assume that the pattern presents rigid rotation, and we use the determination by \citet{Dias_Lepine2005} of $\Omega_{p}=25$ km s$^{-1}$ kpc$^{-1}$, since it is one of the most direct measurements available, as discussed in $\S$~\ref{intro}. This pattern speed, together with the rotation curve from Eq.~\ref{eq:v_L2011}, corresponds to a corotation radius $R_{cr}=8.54$ kpc (indicated by the vertical dash-dotted line in Fig.~\ref{fig:rotcurve}). This value is very close to the upper limit measured by \citet{Dias_Lepine2005} of about 8.55 kpc (for $R_{0}=7.5$ kpc). This radius corresponds to the rotation curve which is flat beyond the solar circle. Adding the Gaussian minimum of Eq.~\ref{eq:funct_mrc} to the curve from Eq.~\ref{eq:v_L2011}, the corotation changes to $R_{cr}=8.1$ kpc (indicated by the vertical dashed line in Fig.~\ref{fig:rotcurve}). It is interesting to note that, recently, \citet{Bobylev_Bajkova2012} estimated the pattern speed of the Milky Way by the analysis of the phase of the spiral arms relative to the Sun for groups of Galactic Cepheids with different mean ages. They also found a pattern speed $\Omega_{p}=25$ km s$^{-1}$ kpc$^{-1}$, for a four-armed spiral model and angular rotation velocity at the solar orbit radius $\Omega_{0}=30$ km s$^{-1}$ kpc$^{-1}$. Our model Sp5 is an exception, as it includes two pattern speeds; the reason for that choice is explained in $\S$~\ref{other_spiral_models}.

The ratio of the field strength $f_{r0}$ (as named by \citealt{Yuan1969}), which is the ratio of the radial component of the spiral field to the axisymmetric field at the solar radius, is given by:

\begin{equation}
\label{eq:fr0}
f_{r0}=\frac{A_{sp_{0}}m}{R^{2}_{0}\Omega^{2}_{0}\tan i}
\end{equation}
where $A_{sp_{0}}$ and $\Omega_{0}$ are the amplitude of spiral perturbation and angular rotation velocity at the solar radius, respectively. \citet{Yuan1969} and \citet{Lin_Yuan_Shu1969} proposed a ratio $f_{r0}=0.05$ for a pattern speed $\Omega_{p}=13.5$ km s$^{-1}$ kpc$^{-1}$. \citet{Roberts_Hausman1984} studied the formation mechanisms of spiral structure and adopted a model with the force ratio ranging from 0.05 to 0.1 in the region of the disk between 5 and 10 kpc. Based on the velocity field of a sample of Cepheids, \citet{Mishurov1979} derived $f_{r0}\approx 0.125$. \citet{Patsis1991} analysed self-consistent models for a sample of 12 normal spiral galaxies and presented a positive correlation between pitch angle and the ratio $f_{r}$; the force ratio increases with the pitch angle, from the early type Sa to the late types Sb and Sc galaxies. According to their figure 15, a galaxy with pitch angle $i\approx 12^{\circ}$ admits a relative force perturbation $f_{r}$ ranging from $\sim$ 0.03 to 0.08. In our experiments, we decided to perform tests adopting $f_{r0}$ in the range 0.05 to 0.15 for the spiral potential with cosine profile from Eq.~\ref{eq:potspiral_real}. For the spiral potential with Gaussian profile, we used the same ratio as given by JLBB, which lies in the interval $0.03-0.06$.

We adopt the same radial profile for the amplitude of the spiral perturbation as given by \citet{Contopoulos_Grosbol1986}, and also by JLBB:

\begin{equation}
\label{eq:amplitude_spiral}
A_{sp}(r)=\mathcal{A}_{\mathrm{sp}}\,r\,e^{-r/r_{sp}},
\end{equation}
where $r_{sp}$ is the scalelength of the spiral arms. In our models, we adopt the same scalelength for the arms and the disk, $r_{sp}=r_{d}=2.5$ kpc. With the parameters of the spirals used in our models, we have values of the amplitude $\mathcal{A}_{\mathrm{sp}}$ ranging from $\sim$ 300 to 1800 km$^{2}$ s$^{-2}$ kpc$^{-1}$. This range of amplitudes is similar to that used by \citet{Antoja2011} in their TWA model of spiral arms, as well as the range adopted by \citet[hereafter MF10]{Minchev_Famaey2010} in their study of radial migration in galactic disks. 

Another important property of the spiral structure is the surface density contrast, defined by the ratio $\delta \Sigma/\Sigma$, where $\delta \Sigma$ is the amplitude of the perturbed surface density, related through Poisson's equation to the amplitude of the spiral potential, and $\Sigma$ is the surface density of the axisymmetrically distributed mass in the disk. 
Based on the recent literature, \citet{Antoja2011} found that estimates of the Galactic surface density contrast lie in the interval 0.1 - 0.23. Our models comprise a similar range of values for this quantity, considering some values for the local surface density $\Sigma_{0}$ estimated in the literature. It must be noted that such values refer to the density contrast of the stellar component; the equivalent quantity for the gas component may be larger, since the relative amplitude of the density wave is inversely proportional to the velocity dispersion of the class of object in the Galactic mid-plane.

\subsection{Resonances}
\label{resonances}

\begin{figure}
 \includegraphics[scale=0.48]{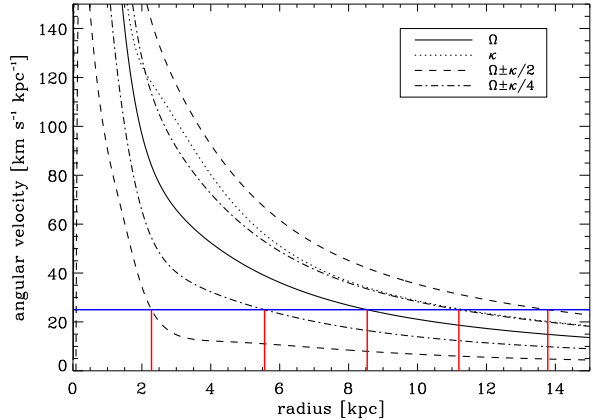}
 \caption{Curves of angular rotation velocity $\Omega$ (solid) and epicycle frequency $\kappa$ (dotted) versus galactic radius. The curves $\Omega\pm \kappa/2$ and $\Omega\pm \kappa/4$ are shown as dashed and dash-dotted lines, respectively. The positions of the main resonances of the spiral structure are indicated by the red vertical lines; the pattern speed of the spiral arms $\Omega_{p}=25$ km s$^{-1}$ kpc$^{-1}$ is indicated by the blue horizontal line.}
\label{fig:rotcurve_angCG86}
\end{figure}

Figure~\ref{fig:rotcurve_angCG86} shows the curves of angular rotation velocity $\Omega=V_{c}/r$, derived from Eq.~\ref{eq:v_L2011} (solid line), and the epicycle frequency $\kappa$ for nearly circular orbits (dotted line). The positions of the Lindblad resonances (LRs) and the corotation of the spiral structure are indicated by the red vertical lines. The LRs occur when $\Omega_{p}=\Omega\pm\kappa/m$; the inner Lindblad resonance (ILR) and the outer Lindblad resonance (OLR) corresponds to the negative and positive signs, respectively. To avoid confusion about the terminology of the Lindblad resonances for different pattern multiplicity, we adopt the same notation used by MF10: 
the 2:1 ILR/OLR and the 4:1 ILR/OLR are referred to for both two-armed and four-armed spiral structure, irrespective of being a first or second-order resonance. 



\section[]{Theoretical background for the radial density minimum}
\label{effec_pot}

\subsection{Effective potential}
\label{poteffect}

We study the stellar orbits under the Galactic gravitational potential in a reference frame corotating with the spiral perturbation, at a constant angular speed $\Omega_{p}$ about the {\it z\/}-axis. 
We restrict to the study of the orbits in the plane of the Galaxy ($z=0$), taking the zero-thickness disk approximation. The polar coordinates of this reference frame are ($r;\theta$), and the relation between the azimuthal coordinates of the inertial and rotating frames is: $\theta=\theta'-\Omega_{p}t$. The Hamiltonian of a star in this rotating frame can be written as:

\begin{equation}
\label{eq:Hamiltonian}
H(r,\theta,\dot{r},\dot{\theta})=\frac{1}{2}\left[\dot{r}^{2}+r^{2}\dot{\theta}^{2}\right]+\Phi_{\mathrm{eff}}(r,\theta),
\end{equation}  
where the effective potential $\Phi_{\mathrm{eff}}$ is given by:

\begin{equation}
\label{eq:phieff}
\Phi_{\mathrm{eff}}(r,\theta)=\Phi_{ax}(r)+\Phi_{sp}(r,\theta)-\frac{1}{2}\Omega^{2}_{p}r^{2}.
\end{equation}



\subsection{Lagrangian points and stability}
\label{Lpoints_stability}

\begin{figure}
	\includegraphics[scale=0.60]{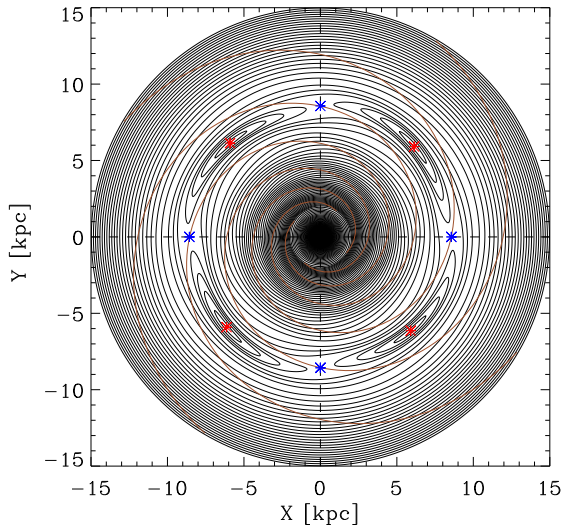}
	\caption{Contours of constant effective potential $\Phi_{\mathrm{eff}}$. The parameters of the spiral potential are: $m=4$, $i=-12^{\circ}$, $f_{r0}=0.05$. Blue asterisks show the positions of the Lagrangian points $L_1$ and $L_2$; red asterisks show the positions of the points $L_4$ and $L_5$. The brown spiral curves correspond to the locus of the minimum of the spiral potential.}
	\label{fig:Hillcurves_poteffect}
	\end{figure}

The term inside the brackets in Eq.~\ref{eq:Hamiltonian} is the velocity of the star in the rotating frame: $\mathit{v}^{2}=\dot{r}^{2}+r^{2}\dot{\theta}^{2}$. Rewriting Eq.~\ref{eq:Hamiltonian} as $H=\frac{1}{2}\mathit{v}^{2}+\Phi_{\mathrm{eff}}$, we see that the curves $H=\Phi_{\mathrm{eff}}$ define the so-called zero-velocity curves (or Hill's curves); the star's motion is restricted to the space region where $\Phi_{\mathrm{eff}}<H$, since $\mathit{v}^{2}$ must be positive. Figure~\ref{fig:Hillcurves_poteffect} shows the zero-velocity curves (contours of constant $\Phi_{\mathrm{eff}}$ for corresponding values of $H$) for the effective potential composed by the axisymmetric component from Eq.~\ref{eq:Phiax} and the spiral potential of the {\it basic model\/} ($m=4$ arms and pitch angle $i=-12^{\circ}$) with $f_{r0}=0.05$ and the amplitude with cosine profile (Eq.~\ref{eq:potspiral_real}). The stationary points in this reference frame are the often called Lagrange points, where the componentes of the gradient of the effective potential mutually vanish. The Lagrangian points $L_1$ and $L_2$ (blue asterisks in Fig.~\ref{fig:Hillcurves_poteffect}) are saddle points of $\Phi_{\mathrm{eff}}$, representing unstable equilibrium points; the points $L_4$ and $L_5$ (red asterisks in Fig.~\ref{fig:Hillcurves_poteffect}) are maxima of $\Phi_{\mathrm{eff}}$, and are stable equilibrium points for certain values of the perturbation amplitude. The Lagrangian point $L_3$ (not shown in Fig.~\ref{fig:Hillcurves_poteffect}) corresponds to the minimum of the effective potential at the center of the frame.
Following \citet[$\S$~II]{Contopoulos1973}, the calculated radii of the Lagrangian points are displaced from the corotation circle $r=R_{cr}$ at most 0.05 kpc, considering the largest value of the amplitude of perturbation of our models. Thus we consider that these points are located on the corotation circle, at first approximation.

We follow the notation used by \citet{Fux2001} to describe the stellar orbits according to the values of the Hamiltonian associated with stars corotating at the points $L_{1/2}$ as $H_{12}\equiv\Phi_{\mathrm{eff}}(L_{1/2})$, and at the points $L_{4/5}$ as $H_{45}\equiv\Phi_{\mathrm{eff}}(L_{4/5})$. Stars with energy $H<H_{12}$ do not cross the $H_{12}$ contour and have their orbits confined in regions inside or outside corotation, with trajectories around the galactic center. Stars with energy in the range $H_{12}<H<H_{45}$ can cross the corotation barrier near the points $L_{1/2}$ and explore both sides of corotation, except a small region around $L_{4/5}$ (\citealt{Fux2001}). In the condition of stable points $L_{4/5}$, the Hill's curves are closed oval curves around the maxima of the potential at $L_{4/5}$ (see Fig.~\ref{fig:Hillcurves_poteffect}), as in the restricted three-body problem (\citealt{Contopoulos1973}). Such librational motions are referred to in the literature as `horseshoe orbits' (e.g. \citealt{Goldreich_Tremaine1982}, \citealt{Sellwood_Binney2002}). \citet{Contopoulos1973} also found the existence of two periods of oscillation around $L_{4/5}$: a short period related to the epicyclic period; and a long period of the order of 1 Gyr associated with librations of the epicyclic center around $L_{4/5}$.
For stellar orbits with energy $H>H_{45}$ there are no forbidden regions in the galactic plane. However, as shown by \citet{Contopoulos1973}, there is a `third' integral of motion, besides the Jacobi integral, that delimits the boundaries of the orbits near the points $L_{4/5}$; such orbits fill rings around $L_{4/5}$ or describe the so-called {\it banana-type\/} orbits (e.g. \citealt{Contopoulos1973}, \citealt{Barbanis1970}, \citealt{Barbanis1976}).

The stability of the points $L_{4/5}$, which is the required condition for the existence of trapped orbits around such points, can be found by the analysis of the solutions of the equations of motion of nearby orbits. Following again \citet[$\S$~II]{Contopoulos1973}, and \citet[$\S$~3.3.2]{Binney_Tremaine2008}, we calculate the maximum amplitude of perturbation that is allowed for the stability condition of the points $L_{4/5}$. 
For the spiral parameters, rotation curve, and the spiral pattern speed $\Omega_{p}$ of our models, we find that the points $L_{4/5}$ are stable for amplitudes of the spiral perturbation $\mathcal{A}_{\mathrm{sp}}<\mathcal{A}_{\mathrm{sp_{max}}}\approx2400$ km$^{2}$ s$^{-2}$ kpc$^{-1}$. Since the amplitudes $\mathcal{A}_{\mathrm{sp}}$ of the models lie in the range $300-1800$ km$^{2}$ s$^{-2}$ kpc$^{-1}$ ($\S$~\ref{pot_non-axis}), we conclude that all our models present stable equilibrium points $L_{4/5}$.


\subsection{Jacobi integral and angular momentum variations in the galactic disk}
\label{JacobiInteg_dL}

We can rewrite the Hamiltonian of Eq.~\ref{eq:Hamiltonian}, with the use of Eq.~\ref{eq:phieff}, as: $H=\frac{1}{2}\left[\dot{r}^{2}+\frac{L^{2}}{r^{2}}\right]+\Phi_{ax}+\Phi_{sp}-\Omega_{p}L$, where $L$ is the specific angular momentum of the star (in the $z$ direction) in the inertial frame. This can be rewritten again as:

\begin{equation}
\label{eq:Jacobi_Integral}
H=E_{J}=E-\Omega_{p}L,
\end{equation}
where $E$ is the total specific energy of the star in the inertial frame. The quantity $E_{J}$, known as the Jacobi integral, is the classical integral of motion in the rotating frame. $E$ and $L$ are not conserved individually, and variations in these quantities are related by:

\begin{equation}
\frac{\ud E}{\ud t}=\Omega_{p}\frac{\ud L}{\ud t}.
\end{equation}
Next, we highlight some results of the theory of energy and angular momentum transfer at resonances developed by \citet[hereafter LBK]{Lynden-Bell_Kalnajs1972}, and the extension of this theory proposed by \citet[hereafter Z96, Z98 and Z99, respectively]{Zhang1996,Zhang1998,Zhang1999} to account for the secular redistribution of the disk surface density in spiral galaxies; these results will be useful in the analysis of the numerical simulations of $\S$~\ref{result_analys}.

LBK showed that for stars on circular orbits, which have the minimum orbital energy, a change in the specific energy is related to a change in the specific angular momentum through: $\partial E/\partial L=\Omega(r)$. The process of steady angular momentum transfer between the spiral density and a star in resonant motion with the perturbation is understood in the following way: 
the loss (gain) of angular momentum $\Delta L$ by a star at the inner (outer) Lindblad resonance is accompanied by the loss (gain) of energy $\Omega_{p}\Delta L$; at the same time, the change in orbital energy, relative to the circular motion, is $\Omega\Delta L$, which in both cases is lesser than $\Omega_{p}\Delta L$ in relative values. Thus the amount of energy $\Delta E_{r}$:

\begin{equation}
\Delta E_{r}=(\Omega_{p}-\Omega)\Delta L
\end{equation}
acquired by the star at the ILR or OLR appears as non-circular motion. It can be noticed that the above expression always has a positive sign, since the terms $\Omega_{p}-\Omega$ and $\Delta L$ are both negative at the ILR and positive at the OLR. It is easy to see that at corotation, where $\Omega=\Omega_{p}$, a star gains or loses angular momentum $\Delta L$ without changing the energy associated with non-circular motion.
Close to the corotation resonance, LBK predicted that: stars just inside corotation, with angular velocities faster than the wave, slow down when feeling the forward pull of a spiral arm; 
on average, they will absorb angular momentum from the wave. On the other hand, stars just outside corotation, with angular velocities slower than the wave, accelerate when held back by the arm; 
there will be an excess of stars giving angular momentum to the wave. This is the `donkey behaviour', characterized by the authors, coupled with an analog of Landau damping for stars corotating with the perturbation. Considering the decrease of the disk surface density with galactic radius, there are more stars just inside than outside corotation, so the net effect at this resonance is the exchange of angular momentum from the wave to the stars. This mechanism naturally explains the formation of the horseshoe orbits around the librational points $L_{4/5}$: the gain of angular momentum by a star initially inside and close to the corotation radius causes the increase of its orbital radius, and eventually, if $r>R_{cr}$ ($\Omega<\Omega_{p}$), its motion relative to the wave is reversed. Outside corotation, the interaction with the subsequent wave causes the loss of angular momentum by the star and its excursion to a smaller radius ($r<R_{cr}$), and thus returning to the advanced drift motion relative to the wave. For a complete scenario, we have to add the epicyclic motion of the star to this libration around $L_{4/5}$.

In a series of papers, Z96, Z98 and Z99 analysed the mechanism of angular momentum transport in galactic disks by means of a collective dissipation process resultant from the long-range nature of the gravitational interaction between the stars and the spiral arms. The author claims the presence of spiral gravitational shocks that lead to a series of small-angle scatterings when a star crosses a spiral arm in the non-linear regime of the spiral mode. The gravitational instability at the spiral arms is a result of the relative azimuthal phase shift between the potential and density spirals, which is, in turn, a natural property of the spiral waves related through the Poisson equation.

The phase shift indicates the presence of a torque applied by the spiral potential on the spiral density and a consequent transfer of energy and angular momentum between the disk matter and the spiral density wave. In addition, the opposite signs of the spiral potential/density phase shift on both sides of corotation are responsible for the outward transfer of angular momentum in the galactic disk: the disk stars inside corotation lose energy and angular momentum to the wave, leading to a secular decrease of their mean orbital radius, whilst the disk stars outside corotation gain energy and angular momentum from the wave, with the secular increase of their mean orbital radius. In this sense, the work by Zhang complements the work from Lynden-Bell \& Kalnajs regarding the prediction of a secular redistribution of matter, energy and angular momentum between the disk stars and the spiral density wave in regions of the disk not restricted to the vicinity of the wave-particle resonances. 
The evolution of the disk surface density 
determined by the flux of matter in opposite directions on both sides of corotation, 
is the phenomenon that we associate with the formation of a minimum in the stellar density at the corotation radius, which is the central subject addressed in the current paper. 

In $\S$~\ref{rad_distrib}, we compare the distribution of variation of angular momentum in the Galactic disk, resultant from the numerical integration of test particle orbits in the Galactic potential, with the theoretical predictions summarized above. We show that even a non-self-consistent calculation of the response of stellar orbits to the applied smooth axisymmetric and spiral potentials is able to account for the secular changes of the mean orbital radii. Quantitatively, these secular changes agree with the theoretical predictions, being directly related to the angular momentum transfer induced by the phase shift between the spiral potential and the response density.



\section[]{Numerical experiments predicting a density minimum}
\label{num_exp}

Many studies have been devoted to the corotation resonance effects on the stellar orbits, using either analytical or numerical methods, e.g. \citet{Barbanis1970,Barbanis1976}, \citet{Contopoulos1973}, \citet{Mennessier_Martinet1978,Mennessier_Martinet1979}, \citet{Morozov_Shukhman1980}, \citet{Palous1980}, \citet{Bertin_Haass1982}, \citet{Lepine2003}. In this paper, we prefer to study the non-linear interaction of stars with the corotation resonance based on the numerical experiments approach.

We perform test-particle simulations to study the dynamical evolution of a two-dimensional stellar disk in the mid-plane of the Galaxy. We follow the trajectories of non-self-gravitating particles in a gravitational potential initially composed by the background axisymmetric component, in which non-axisymmetric perturbations are grown with time. The orbits are integrated in a reference frame where the spiral arms are stationary, ensuring the conservation of the Jacobi integral, as described in $\S$~\ref{effec_pot}. With the Hamilton's equations applied to the Hamiltonian from Eq.~\ref{eq:Hamiltonian}, and in the reference frame corotating with the perturbation, we have the following equations of motion:

\begin{eqnarray}
\label{eq:eq_motion}
\ddot{r}&=&-\frac{\partial \Phi_{ax}}{\partial r}-\frac{\partial \Phi_{sp}}{\partial r}+r\dot{\theta}^{2}+2\Omega_{p}r\dot{\theta}+\Omega^{2}_{p}r\nonumber \\
\ddot{\theta}&=&-\frac{1}{r^{2}}\frac{\partial \Phi_{sp}}{\partial \theta}-\frac{2}{r}\dot{r}\dot{\theta}-\frac{2}{r}\Omega_{p}\dot{r}.
\end{eqnarray}
The potentials of the axisymmetric component $\Phi_{ax}$ and the spiral perturbation $\Phi_{sp}$ 
are given by Eqs.~\ref{eq:Phiax} and \ref{eq:potspiral_real}/\ref{eq:potspiral_gauss}, 
respectively. The total number of particles used in each simulation is $\sim$ 10$^{5}$ particles. The total time of integration is 5 Gyr for most of the simulations, which corresponds to approximately 20 revolutions of the spiral pattern in the inertial reference frame (for a pattern speed $\Omega_{p}=25$ km s$^{-1}$ kpc$^{-1}$). The numerical integrations of the Eqs.~\ref{eq:eq_motion} are performed by means of a fifth-order Runge-Kutta integration scheme, with a typical time step of 0.5 Myr. We checked the conservation of the Jacobi integral of the particles at the end of one orbital period and we found a typical deviation of this quantity on the order of $\left|\Delta E_{J}/E_{J_{0}}\right|\approx 10^{-9}$ (in the cases when only one spiral mode is present).

\subsection{Initial conditions}
\label{init_cond}

The Galactic disk is simulated by choosing randomly the positions of test-particles with a number density distribution in the azimuthal and radial directions with the form:

\begin{equation}
\label{eq:dens_profile}
N_{*}(r)=2\pi N_{c}\,r\,e^{-r/r_{d}}\,\Delta r,
\end{equation}
where $N_{*}$ is the number of stars in the radius range between $r$ and $r+\Delta r$, $N_{c}$ is the central number of stars, and for the disk scalelength we use $r_{d}=2.5$ kpc (\citealt{Freudenreich1998}). This radial profile resembles the modified exponential profile of the Galactic disk surface density proposed by \citet{Lepine_Leroy2000}, which presents a decrease of the density towards the center and an exponential behaviour at larger radii. \citet{Lopez-Corredoira2004} proposed a similar density profile and interpreted the deficit in the central density as a flare in the vertical distribution of stars.

The particles in each radius are assigned with initial circular velocities from the rotation curve of Eq.~\ref{eq:v_L2011}. 
To these initial velocities, the radial $\sigma_{U}$ and azimuthal $\sigma_{V}$ velocity dispersions are added, with a radial profile similar to the one used in Eq.~\ref{eq:dens_profile} and a Gaussian shape at each radius, with the values $\sigma_{U_{0}}=\sigma_{V_{0}}=5$ km s$^{-1}$ at the solar radius and a velocity dispersion peak reaching $\sim 6.5$ km s$^{-1}$. These values are compatible with the amplitudes of the perturbation velocities due to the spiral waves found in the literature (e.g. \citealt{Burton1971}, \citealt{Mishurov1997}, \citealt{Bobylev_Bajkova2010}). They are also similar to the velocity dispersion of the youngest Hipparcos stars (\citealt{Aumer_Binney2009}), as commented by \citet{Antoja2011}.

The spiral perturbation is switched on at the time $t\approx 0.5$ Gyr, reaching its maximum amplitude at $t\approx 2$ Gyr and remaining constant after this time. The increase of the amplitude of the spiral perturbation with time is given by the following function:

\begin{equation}
\label{eq:amp_spiral_growth}
A_{sp}(t)=A_{sp_{max}}\frac{1}{2}\left\{1+\tanh[\,3(t-1.1)\,]\right\},
\end{equation}
with $t$ in Gigayears. This function ensures a smooth transition from the unperturbed initial disk to the perturbed state, as well as a regime of adiabatic growth of the perturbation, since the growth time is larger than the revolution time of the spirals. In this paper, we only consider the time variation of the amplitude of the spirals; the pitch angle and pattern speed are kept constant during the integrations.



\section[]{Results and analysis}
\label{result_analys}

We initially perform the analysis of the results based on the outcome of the simulation that employs the non-axisymmetric model Sp1. This model is composed by our {\it basic model\/} of the spiral geometry ($m=4$ arms and pitch angle $i=-12^{\circ}$), with a ratio of field strength $f_{r0}=0.075$ and a cosine profile for the amplitude of the spirals. The cosine profile is chosen only to make easier the comparison with other studies that employ the same profile for the spiral potential. We will see that the minimum of stellar density at corotation is also produced by the spiral potential with Gaussian profile. This turns to be an important result: the formation of the minimum density at corotation does not have a strong dependence on the azimuthal shape of the spiral arms; its strongest dependence is on the potential/density phase shift, as we will discuss later. The results of the simulations using the other non-axisymmetric models of Table~\ref{tab:models_non-axis} are discussed in $\S$~\ref{other_spiral_models}. Hereinafter, we refer to the test-particles as stars, since this is the role they play in the simulations.

\subsection{Radial distributions}
\label{rad_distrib}

\begin{figure}
	\includegraphics[scale=0.40]{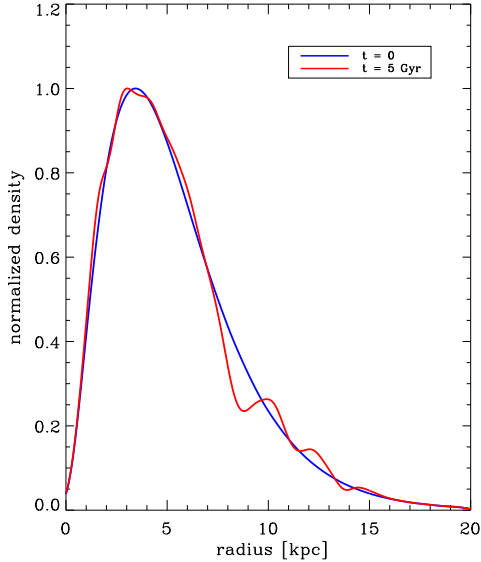}
	\caption{Normalized density distribution of stars in the galactic disk. The curves correspond to the initial distribution at $t=0$ (blue curve), and the azimuthally averaged final distribution at $t=5$ Gyr (red curve). The distributions are the results of a simulation run using the spiral perturbation model Sp1.}
	\label{fig:density_t5Gyr}
	\end{figure}

In Fig.~\ref{fig:density_t5Gyr}, we show the initial and final normalized density distributions of stars in the galactic disk, as blue and red curves, respectively. The final distribution is taken at the end of the simulation, at $t=5$ Gyr. Apart some general features, it can be seen that the largest deviation from the initial distribution occurs at the corotation radius, $R_{cr}=8.54$ kpc. 
A kernel density estimation technique was applied to the distributions, giving the smoothed curves shown in Fig.~\ref{fig:density_t5Gyr} (\citealt{Silverman1986,Bowman_Azzalini1997}).
 

	
\begin{figure}
	\includegraphics[scale=0.45]{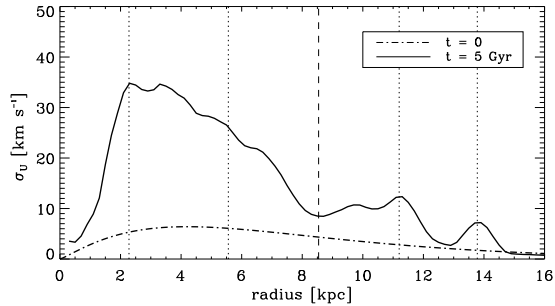}
	\caption{Radial velocity dispersion as a function of radius for the initial unperturbed disk (dash-dotted curve) and for the final state at the end of the simulation (solid curve). The distributions are the results of a simulation run using the spiral perturbation model Sp1. The vertical dashed line indicates the corotation radius and the vertical dotted lines indicate the inner and outer 2:1 and 4:1 Lindblad resonances.}
	\label{fig:sigVr_simulat}
	\end{figure}

Figure~\ref{fig:sigVr_simulat} shows the distribution of the radial velocity dispersion as a function of radius, at the beginning (dash-dotted curve) and at the end of the simulation (solid curve). The peaks in the final distribution of $\sigma_{\scriptscriptstyle U}$ are related to the Lindblad resonances. It can also be noted a local minimum in the final distribution related to the corotation radius, which is in agreement with the fact that changes in angular momentum at corotation do not produce significant dynamical heating of the stellar disk.

\begin{figure*}
	\includegraphics[scale=0.47]{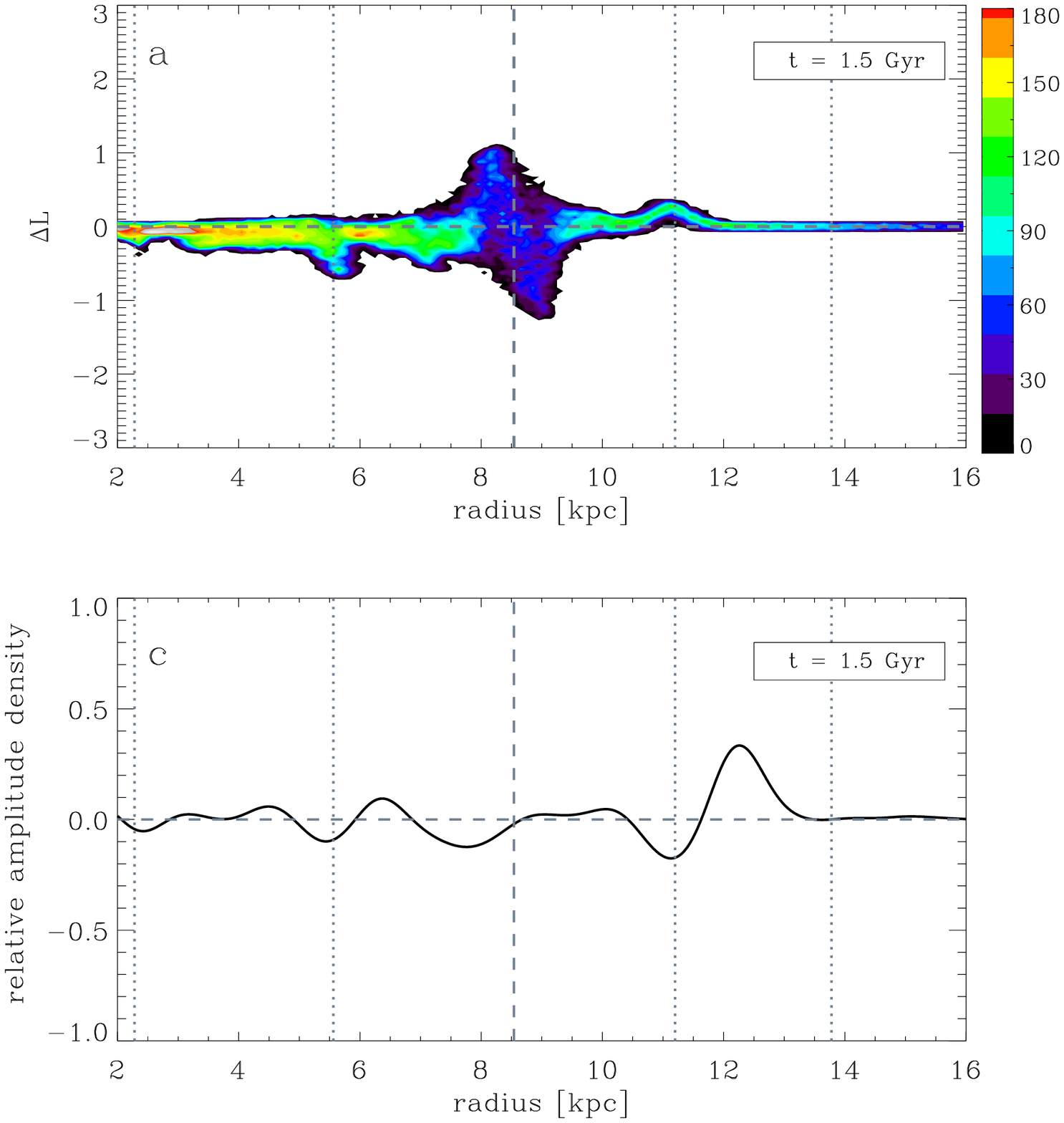}
	\includegraphics[scale=0.47]{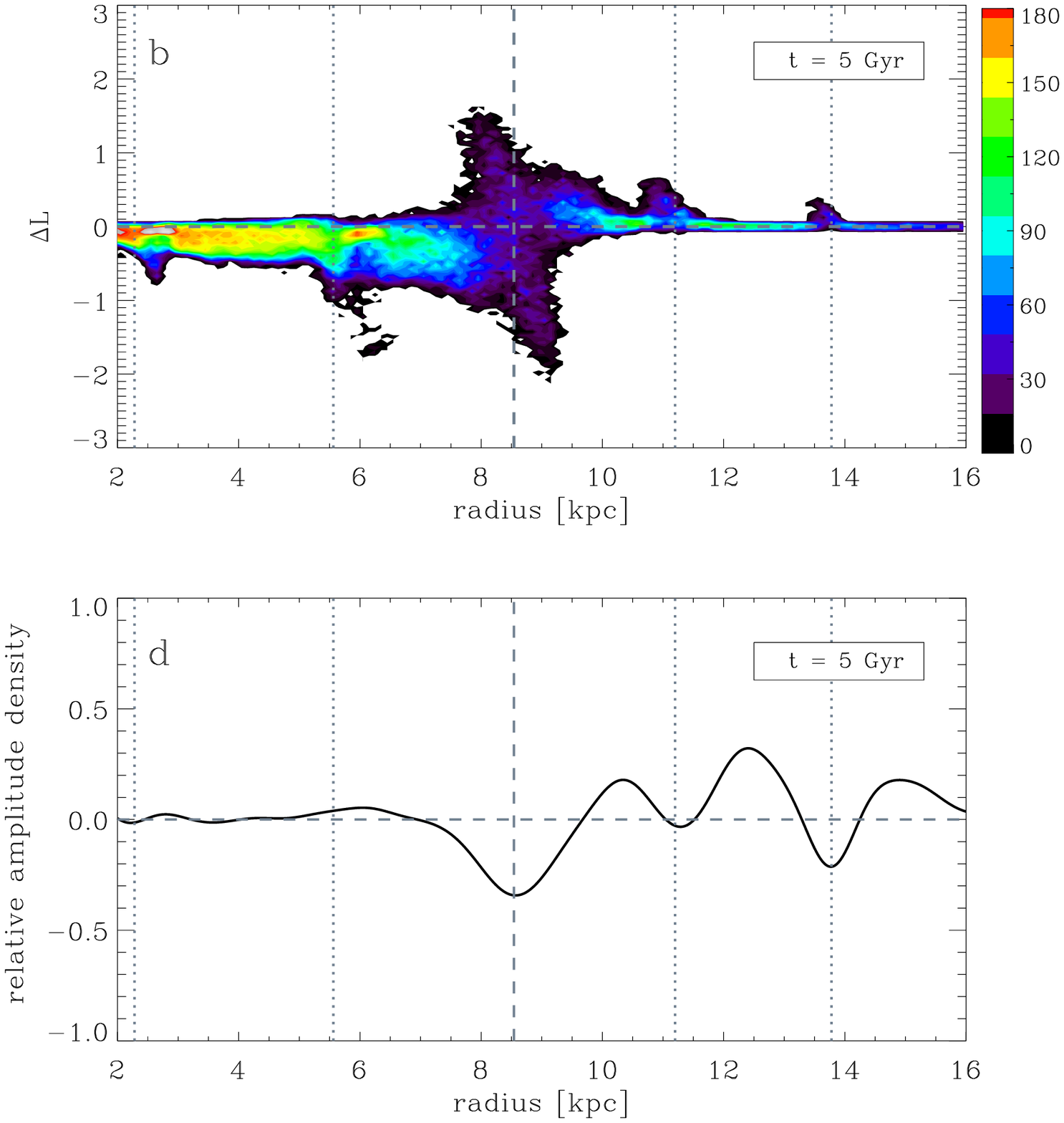}
	\caption{{\bf Top row:} Changes in angular momentum as a function of the initial galactic radius. The {\it y\/}-axis shows $\Delta L$ in units of the initial circular velocity $V_{c}$ of the stars. The color scale is a measure of the density of stars in each bin of $\Delta L$ and $r$. {\bf Bottom row:} Amplitudes of the relative changes between the initial and final (at a time step $t$) density distributions of stars in the galactic disk. The left and right panels of each row show the distributions at the time steps $t=1.5$ and 5 Gyr, respectively. {\bf Vertical lines:} The corotation radius is marked by the dashed lines; the 2:1 ILR/OLR and 4:1 ILR/OLR are indicated by the dotted lines. The distributions are the results of a simulation run using the spiral perturbation model Sp1.}
	\label{fig:distr_dL_amprel}
	\end{figure*}

We estimate the changes in angular momentum $\Delta L$ across the galactic disk by calculating the initial $L_{i}$ and final $L_{f}$ angular momenta of each star. The final angular momenta are averaged over the last two revolutions of each individual stellar orbit. The top row of Fig.~\ref{fig:distr_dL_amprel} shows $\Delta L$ as a function of the initial galactic radius, at the time step $t=1.5$ Gyr (left panel), and at $t=5$ Gyr (right panel). The color scale corresponds to the density of stars in the $\Delta L-r$ plane. The {\it y\/}-axis of each panel actually shows the angular momentum variation $\Delta L$ divided by the initial circular velocity $V_{c}$ of the stars, which approximately gives the magnitude of the changes in radius (hereinafter, although not explicitly, the values for $\Delta L$ are given in units of kiloparsecs). The bottom row shows the relative amplitude of variation of stellar density as a function of radius; in other words, it is the difference between the initial and final density distributions (shown in Fig.~\ref{fig:density_t5Gyr}), divided by the initial one. The left and right panels also show the distributions at the time steps $t=1.5$ and 5 Gyr. 
In each panel of Fig.~\ref{fig:distr_dL_amprel}, the vertical dashed line indicates the corotation radius of the galactic model; the vertical dotted lines indicate the radii of the inner and outer 2:1 and 4:1 Lindblad resonances. 
	
As already noted by other authors (e.g. \citealt{Sellwood_Binney2002}; MF10; \citealt{Roskar2012}), the largest changes in angular momentum occur near corotation. The shape of $\Delta L$ with a negative slope, on the order of $-2$ with respect to the line $\Delta L=0$, causes some stars to move symmetrically from one to the other side of corotation. The general trend is an increase in $\Delta L$ just inside the corotation radius and a decrease just outside it (\citealt{Minchev2012}). In our experiments, this behaviour is especially verified during the stages of growth of the amplitude of perturbation. Immediately after the time when the spiral perturbation is switched on, at $t\approx 0.5$ Gyr, the angular momentum changes start growing exclusively at the Lindblad resonances and corotation resonance. During the late stages of the perturbation growth, the distribution $\Delta L(r)$ reaches local maxima at those resonances, but some spread in $\Delta L$ along the entire galactic disk can be observed. This can be verified analysing Fig.~\ref{fig:distr_dL_amprel}a: the distribution is recorded at the time step $t=1.5$ Gyr, which corresponds to the time when the spiral perturbation reaches $\sim 92\%$ of its maximum amplitude, according to the function in Eq.~\ref{eq:amp_spiral_growth}. In addition to the aforementioned features, we observe that the signs of $\Delta L$ at the resonances are in agreement with the predictions by LBK for the linear phase of the spiral wave growth, as discussed in $\S$~\ref{JacobiInteg_dL}. Despite the relative spread of $\Delta L$ around corotation, a structure with significant density of stars is clearly visible: regions with large positive and negative $\Delta L$ just inside and outside corotation, respectively, connected by the ridge with negative slope discussed previously.

After the full growth of the amplitude of the spirals, at $t\approx 2$ Gyr, we observe that the changes in $L$ take place over the entire galactic radius, not only at the resonances. A comparison between the Figs.~\ref{fig:distr_dL_amprel}a and~\ref{fig:distr_dL_amprel}b reveals that a large fraction of angular momentum is exchanged even during the stages when the perturbation is fully grown and constant in time. The main variations in $L$ occur until the end of the simulation, shown in the Fig.~\ref{fig:distr_dL_amprel}b; we performed simulations with a total integration time of $t=10$ Gyr and only small further variations in the distribution of $\Delta L(r)$ were observed after the step $t=5$ Gyr. Furthermore, it can also be observed in Fig.~\ref{fig:distr_dL_amprel}b that the sign of $\Delta L(r)$ reverses across corotation: the main changes in $L$ are negative for radii inside corotation and positive for radii outside it. These results are in close agreement with the works from Z96 and Z98, regarding her predictions about the angular momentum exchanges between the disk stars and the spiral wave in the non-linear regime of the wave mode, which is when the spiral mode reaches its quasi-steady state. We shall return to the analysis of the correspondence between our results and the work from Zhang later in this section.  

We now focus on the variations of angular momentum and stellar density that take place near the corotation circle. During the initial growth of the spiral perturbation, the amplitudes of $\Delta L$ increase with the amplitude of the spirals. If we analyse the changes in $L$ in an annulus with width $\sim 0.5$ kpc just inside the corotation radius, from Fig.~\ref{fig:distr_dL_amprel}a we see a larger density of stars presenting positive values for $\Delta L$ rather than negative values. The opposite is seen when we take the annulus just outside corotation. These changes in $L$ are due to the stars moving on horseshoe orbits. As the amplitude of the spirals increases, the process of orbit trapping around the librational points $L_{4/5}$ also increases, enabling the development of large changes in $L$. 
At the full growth of the perturbation ($t\approx 2$ Gyr), the amplitudes of $|\Delta L|$ reach $\approx 1.4$ on both sides of corotation. During the late growth and in the time-independent phase of the amplitude of the spirals, from $t=2$ to 5 Gyr, the development of negative changes in $L$ ($\Delta L<0$) by stars in the annulus inside corotation seems to counterbalance the positive increase of $L$ due to the horseshoe orbits. The opposite process happens just outside corotation. However, from $t=2$ to 5 Gyr, we observe a further increase of the positive peak of $\Delta L$ inside corotation, from $\approx 1.4$ to $\approx 1.7$; and also an absolute increase of the negative peak outside corotation, from $\approx -1.4$ to $\approx -2.0$. This can be explained in the following way: some stars traveling on horseshoe orbits can suffer changes in $L$ with the opposite sign needed for the maintenance of such orbits, being either inside or outside corotation. For example, a star that was initially in a horseshoe orbit and in the outer side of corotation (with energy $H_{12}<H<H_{45}$), loses angular momentum when interacting with the spiral wave and slips to the inner side of the horseshoe orbit ($r<R_{cr}$). Rather than gaining $L$ from the wave in this side of the orbit, the star can lose angular momentum to the wave since at the time-independent phase of the spirals, a large fraction of stars at radii inside corotation start presenting negative values of $\Delta L$. After some revolutions, these losses of angular momentum can eventually cause the `detachment' of the star from the initial horseshoe orbit and lead it to an inner orbit around the galactic center (with energy $H<H_{12}$). A similar picture with the opposite signs of $\Delta L$ can be thought for a star initially in the inner side of the horseshoe orbit; the star will end up in a bounded orbit around the galactic center and in the outer side of corotation. This mechanism explains the further increment on the peaks of $\Delta L$ at the extremities of the ridge with negative slope, during the constant phase of the spiral amplitude. 

\begin{figure*}
	\includegraphics[scale=0.70]{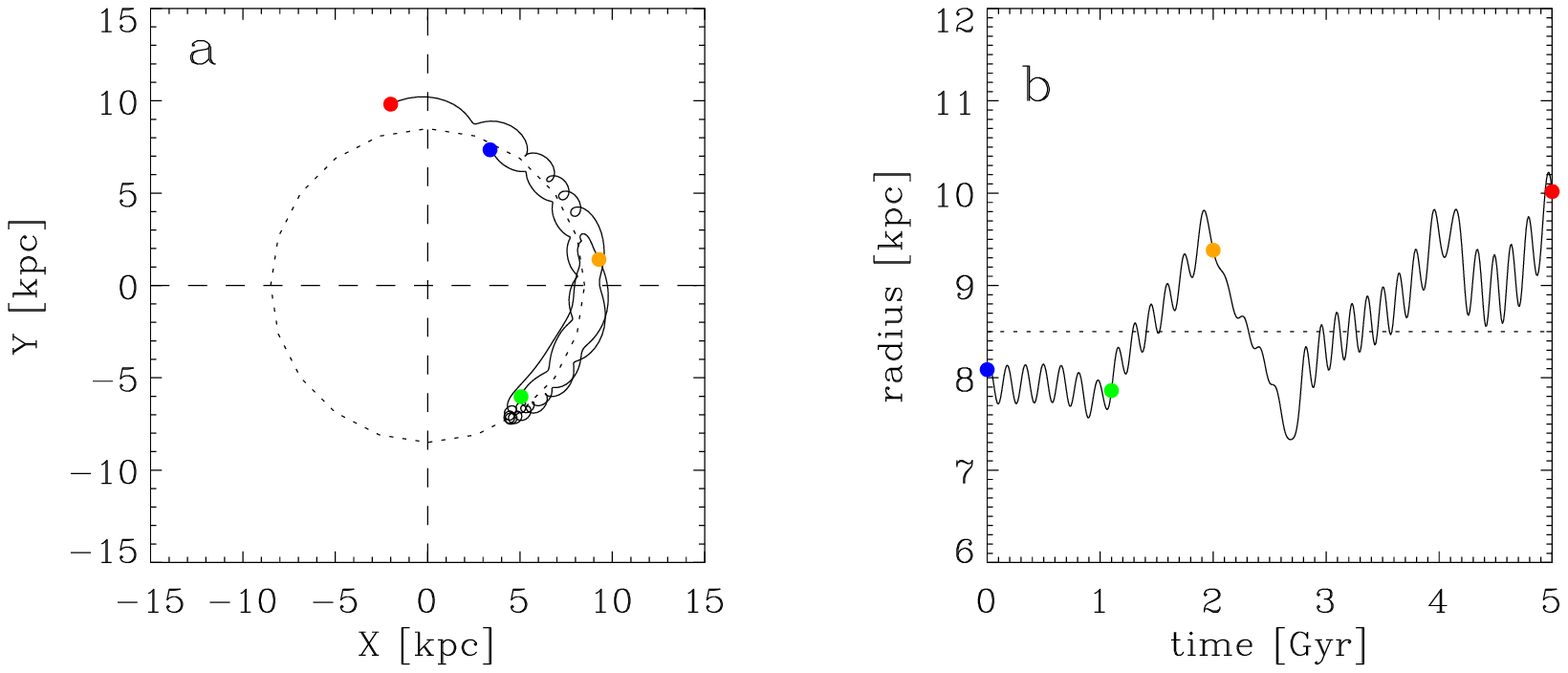}
	\includegraphics[scale=0.70]{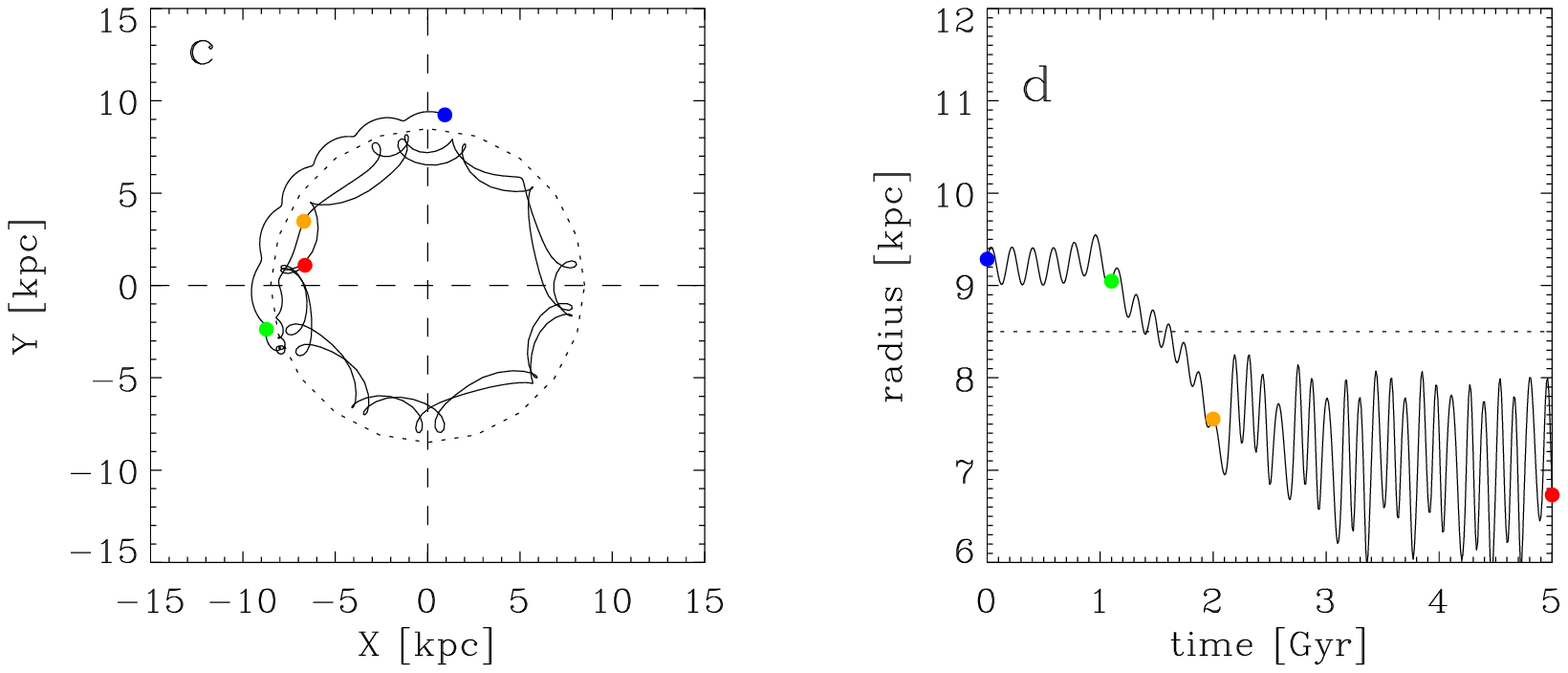}
	\caption{{\bf Left column:} (a) Orbital trajectory in the galactic plane for a star initially inside corotation and (c) initially outside it. The orbits are in the frame corotating with the spiral arms, and the positive direction of rotation is clockwise. The dotted curve indicates the corotation circle. {\bf Right column:} Time evolution of the orbital radius for the orbits shown in the left column. The horizontal dotted line indicates the corotation radius. In each panel, the blue and red dots indicate the initial and final positions of the star, respectively; the green dot indicates the position at $t=1.1$ Gyr, and the orange dot marks the position at $t=2$ Gyr.}
	\label{fig:orbs_near_CR}
	\end{figure*}

In the panels of Fig.~\ref{fig:orbs_near_CR}, we give examples of such scattering process at corotation happening for two stars in our simulations. In Figs.~\ref{fig:orbs_near_CR}a and~\ref{fig:orbs_near_CR}c, we plot the orbits of the stars in a frame corotating with the spirals; the blue and red dots show the inital and final positions, respectively, the green dots show the positions when the spiral amplitude is one half of its maximum (at $t=1.1$ Gyr) and the orange dots mark the positions when the amplitude reaches its maximum (at $t=2$ Gyr). Figures~\ref{fig:orbs_near_CR}b and~\ref{fig:orbs_near_CR}d show the time variation of the radius of each orbit. From Fig.~\ref{fig:orbs_near_CR}a and~\ref{fig:orbs_near_CR}b, we see that the star was initially inside corotation, got trapped in a horseshoe orbit between $t \sim 1$ to 4 Gyr, and then escaped from the trapping zone and ended up in an orbit in the outer side of corotation. Throughout the process, the star was scattered from one to the other side of corotation without significant increasing of the eccentricity of its orbit. In the case of Figs.~\ref{fig:orbs_near_CR}c and~\ref{fig:orbs_near_CR}d, the inverse process occurred: the star was initially outside corotation and ended up in a highly eccentric orbit inside corotation. We see that during the corotation crossing, between $t \sim 1$ - 2 Gyr, no increase in the amplitude of the epicyclic motion had happened, and the further significant increase of this quantity until the end of the simulation may have been caused by large angular momentum exchanges inside corotation.

In the framework of recurrent spiral waves, \citet{Sellwood_Binney2002}, also \citet{Roskar2012}, argue that only transient spirals are capable to promote the excursion of the stars from one to the other side of corotation. According to these authors, the time variation of the perturbation, since its exponential growth to the rapid vanishing, is on the order of one half of the period of the horseshoe orbits, so stars in these orbits are able to cross corotation at most once. In this way, a steady spiral perturbation would not cause scattering across corotation because the stars would be trapped on horseshoe orbits all the time, resulting in no angular momentum exchange near corotation. Our simulations contradict this last hypothesis by the mechanism discussed previously: in a steady spiral perturbation, stars initially on horseshoe orbits, through a process of loss/gain of $L$ inside/outside corotation (contrary to the one needed for the horseshoe orbits), can gradually come out of the trapping process and migrate to orbits with radii inside or outside corotation and around the galactic center. These angular momentum exchanges with opposite signs from the ones shown by stars in horseshoe orbits are what we relate to the mechanism proposed by Z96 responsible for the stabilization of the wave and the redistribution of the stellar surface density along the galactic disk. Therefore, we conclude that a steady spiral wave is able to promote radial migration across corotation; some imprints of this process may be present in the radial distribution of metallicity near the corotation circle, 
a subject that we address in a forthcoming paper.

\begin{figure}
	\includegraphics[scale=0.45]{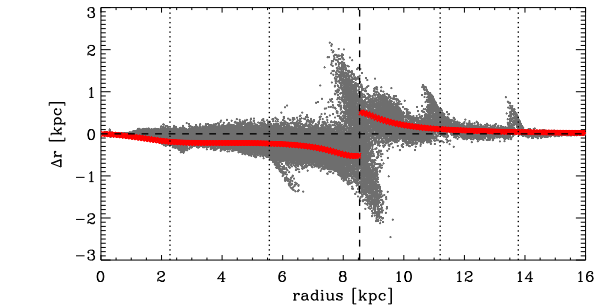}
	\caption{Changes in the mean orbital radius as a function of the initial radius for the stars in the simulation run using the spiral model Sp1. The sequence of red points show the distribution of $\Delta r$ using the analytical expressions from Eqs.~\ref{eq:drmed_dt} and~\ref{eq:phase_shift}. The vertical dashed line indicates the corotation radius and the vertical dotted lines indicate the inner and outer 2:1 and 4:1 Lindblad resonances.}
	\label{fig:deltaR_r}
	\end{figure}

Hereinafter, we refer to the angular momentum exchanges that take place over the entire galactic disk, in the steady state of the spirals, as the {\it secular\/} changes in $L$, to make distinction from the changes in $L$ that occur at the resonances during the linear growth of the spirals. From the top right panel of Fig.~\ref{fig:distr_dL_amprel}, we see that the main secular changes in the angular momentum of the stellar disk are negative inside corotation, and positive outside it. These secular changes in $L$ can also be observed as secular changes in the mean orbital radii of the disk stars, with the same pattern of signs relative to corotation. In Fig.~\ref{fig:deltaR_r}, we plot the changes in the mean orbital radius, $\Delta r=r_{f}-r_{i}$, as a function of the initial galactic radius $r_{i}$ of the stars; the final radii $r_{f}$ are the mean radii of the last two revolutions of each individual stellar orbit.
Z98 obtained an expression for the rate of secular change of the mean orbital radius $r_{*}$ of a star due to the interaction with an open spiral pattern in the quasi-steady state, which is written as:

\begin{equation}
\frac{\ud r_{*}}{\ud t}=-\frac{1}{2}F^{2}V_{c}\;\tan i\;\sin(m\phi_{0}),
\label{eq:drmed_dt}
\end{equation}
where $F^{2}\equiv F_{\Sigma}F_{\nu}$. Relating to our parameters, $F_{\Sigma}$ is the surface density contrast $\delta \Sigma/\Sigma$, and $F_{\nu}$ is the ratio of field strength $f_{r}$; $V_{c}$ is the circular velocity, $i$ is the pitch angle, $m$ is the number of arms, and $\phi_{0}$ is the phase shift between the potential and density spirals. For the phase shift, Z96 gives an expression based on the solutions of the linearized Eulerian equations of motion for an orbit in a frame corotating with the spiral pattern (see the Appendix B of Z96):

\begin{equation}
\phi_{0}(r)=\frac{1}{m}\tan^{-1}\left[\frac{-\frac{\ud A_{sp}}{\ud r}-\frac{2\Omega A_{sp}}{r(\Omega-\Omega_{p})}}{A_{sp}\,k}\right],
\label{eq:phase_shift}
\end{equation}
where $k=\frac{m}{r\tan i}$ is the radial wavenumber and all the other parameters, except $\Omega_{p}$, are functions of the zeroth-order radius of the unperturbed orbit. Inserting $\phi_{0}$ from Eq.~\ref{eq:phase_shift} into Eq.~\ref{eq:drmed_dt}, with the parameters of the spiral model Sp1 and the rotation curve of Eq.~\ref{eq:v_L2011}, and considering a constant density contrast of $\delta \Sigma/\Sigma=0.15$, and integrating Eq.~\ref{eq:drmed_dt} iteratively over the radius variation and over a time interval $\Delta t$, we can estimate the magnitude of the secular changes $\Delta r$ along the galactic disk. We do this calculation using a time interval of 3 Gyr, which is the interval of the time-independent phase of the spirals in our simulations. The derived analytical distribution of $\Delta r$ as a function of initial radius is overlaid in the plot of Fig.~\ref{fig:deltaR_r}, shown as a sequence of red points. 
It can be observed a close agreement between the mean values of the distribution $\Delta r(r)$ in bins of galactic radius, obtained with simulations, and the ones predicted by the analytical expressions from Eqs.~\ref{eq:drmed_dt} and~\ref{eq:phase_shift}. It has to be mentioned, however, that this great concordance comes partly from our controlled manner of imposing a spiral perturbation which ideally reaches a phase of stationary state. As stated before, the orbital change rate in the form expressed by Eq.~\ref{eq:drmed_dt} results from the quasi-steady state constraint of the spiral mode (Z99); however, the spiral properties in Eq.~\ref{eq:drmed_dt} ($F$, $i$ and $\phi_{0}$) are functions of radius $r$ as also of surface density $\Sigma(r)$ and velocity dispersion $\sigma(r)$, thus the secular evolution of the stellar disk properties only guarantees the stationary state of the spirals in a self-consistent scenario. Z99 showed that in cases in which such self-consistency is absent, the spiral wave mode must coevolve with the properties of the stellar disk during the process of stellar orbital redistribution. In any case, if the Milky Way's spiral pattern has experienced such a phase of quasi-stationary state over the last few billion years with its current observed characteristics, our simple simulations should give us, at least, an approximate picture of the redistribution of the stellar surface density along the Galactic disk. 

\subsection{The reliability of the test-particle simulations}
\label{reliable_simulat}

The mechanism of gravitational shocks leading to the dissipation of orbital energy when a star crosses a spiral arm, which is described by Z96 as the responsible for the secular evolution of the disks of spiral galaxies, should suitably be inferred by means of N-body experiments. One problematic issue is that most current N-body simulations fail to reproduce spiral patterns that last for several revolutions in a disk galaxy. The N-body simulation performed by Z96 (a version of the simulation from \citealt{Donner_Thomasson1994}) and that gave support to her theoretical findings, produced a stellar disk in which a two-armed long-lived spiral pattern developed spontaneously. Later on, \citet{Sellwood2011} reproduced the Zhang's N-body disk and in his simulations, what was earlier recognized as a long-lived structure, appears now as a superposition of a few strong patterns with different speeds, each of them having short lifetimes. 

We have shown that our test-particle simulations give reasonable quantitative descriptions of the evolution of the stellar disk, at least in the steady phase of the spirals. The main example of this is the good correspondence between the analytical and the mean distribution of the simulated secular changes of mean orbital radii, shown in Fig.~\ref{fig:deltaR_r}. However, one may question how our simulations with non-self-gravitating particles are able to reproduce the evolution of a system in which physical processes involving gravitational instabilities act to drive the secular evolution of the system basic state. 
The instabilities and scattering processes at the spiral arms, pointed out by Z96, 
are direct consequences of the phase shift between the potential and density spirals. Ultimately, the secular changes in the angular momentum of the stars are driven by the torque applied by the spiral potential on the spiral density, and this is also in turn a consequence of the phase shift. In our simulations, the signs of the changes $\Delta L$ and $\Delta r$ on both sides of corotation (Figs.~\ref{fig:distr_dL_amprel}b and~\ref{fig:deltaR_r}) suggest a similar mechanism of torque being applied on the stellar orbits: a negative torque inside corotation and a positive one outside it. Moreover, this could only be achieved if there was also present in our simulations a phase shift between the minimum of the spiral potential and the maximum of the response spiral density at each radii. We next show that such phase shift indeed seems to operate in our simulations. 

	\begin{figure}
	\includegraphics[scale=0.47]{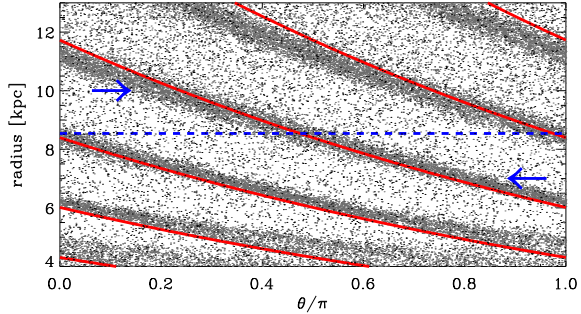}
	\caption{Positions of the particles in a simulation run with minor changes in the initial conditions (see text for details). The red curves indicate the locii of the minimum of the spiral potential; the horizontal dashed blue line marks the corotation radius. The blue arrows point to some regions of maximum response density of particles.}
	\label{fig:poteff_Gauss}
	\end{figure}

In Fig.~\ref{fig:poteff_Gauss} we show the positions of particles, in the $r-\theta$ plane, of a simulation run with minor changes in the initial conditions: 1 - a uniform initial radial density ditribution of particles in the disk; 2 - a spiral perturbation abruptly introduced at the beginning of the simulation and with an amplitude constant in time and independent of the galactic radius. These changes are only intended to reproduce a stronger response density of particles to the spiral perturbation and with similar values along the entire galactic disk, and are not subject to losses of generality. \citet{Mennessier_Martinet1979} and \citet{Palous1980} showed that there is no substantial difference in the stellar orbital response between a slowly and an abruptly introduction of the spiral perturbation. The positions were recorded at the very beginning of the simulation, $t=100$ Myr, when the velocity dispersions are small and the response density is strong. The red curves in Fig.~\ref{fig:poteff_Gauss} indicate the locus of the minimum of the spiral potential; the horizontal dashed blue line indicates the corotation radius. The sense of rotation is of increasing azimuth and the arms represent trailing spirals. It can be observed from this figure the phase shift between the minimum of the spiral potential and the maximum of the response density of particles, as well as the change of the phase shift sign across corotation; for radii inside (outside) corotation, the response density is stronger at azimuths that lead (lag) in phase the corresponding minimum of the spiral potential.

\begin{figure}
	\includegraphics[scale=0.42]{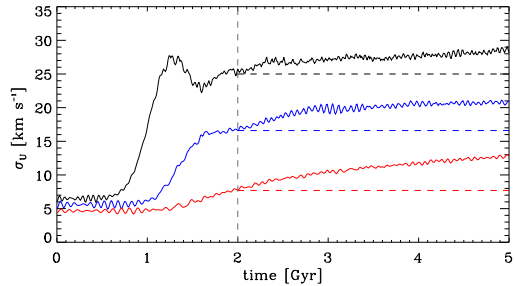}
	\caption{Radial velocity dispersion as a function of time at three different initial radii: $r_{1}=5.8$ kpc (black line), $r_{2}=8.5$ kpc (red line), and $r_{3}=7.3$ kpc (blue line). The curves for $r_{1}$ and $r_{2}$ show the evolution of $\sigma_{U}$ at radii close to the 4:1 ILR and corotation, respectively; the curve for $r_{3}$ shows the evolution of an intermediary region between these two main resonances. The fast rise of $\sigma_{U}$ during the initial growth of the perturbation is depicted in the curves for $r_{1}$ and $r_{3}$. The horizontal dashed lines emphasize the continuation of the disk heating during the steady state of the spirals (after $t=2$ Gyr).}
	\label{fig:sigVr_t}
	\end{figure}

As we discussed in $\S$~\ref{JacobiInteg_dL}, at radii other than corotation, the exchange of angular momentum and energy between a star and the spiral wave is followed by the conversion of part of the stellar orbital energy into heat (i.e., non-circular motion). But now, this heating process occurs at any radii, not only at resonances. With the increase of the amplitude of the epicyclic motion, the phase of the non-circular component of the stellar velocity becomes decorrelated with the phase of the spirals, leading to a saturation of the amplitudes of the changes in energy and angular momentum. In Fig.~\ref{fig:sigVr_t} we show the heating of the galactic disk in terms of the time variation of the radial velocity dispersion, based on the results of the simulation run with the model Sp1. We plot the curves of $\sigma_{U}$ for three different initial radii: $r_{1}=5.8$ kpc, which is close to the 4:1 ILR; $r_{2}=8.5$ kpc, which is very close to corotation; and for an intermediary radius $r_{3}=7.3$ kpc. The time variations of each curve agree with our expectations about the process of angular momentum and energy transfer during the stages of the perturbation growth and its steady phase. This is also, in turn, a proof that our simulations incorporate dissipative effects that lead to the heating of the stellar disk. It has to be noted that the curves $\sigma_{U}(t)$ in Fig.~\ref{fig:sigVr_t} show the evolution of the heating of the stars that were initially at the radii $r_1$, $r_2$ and $r_3$, but not necessarily remained at these radii as the time progresses, since we expect that these stars suffered a decrease in their mean orbital radii. Consequently, with the increasing distance from corotation, a given amount of loss of angular momentum leads to a larger amount of energy converted into non-circular motion. Fig.~\ref{fig:sigVr_t} shows that even after the time when the spiral wave reaches its steady state ($t=2$ Gyr), the heating process continues to operate in the disk at any radii other than corotation (Z99), at a smaller rate than during the phase of the spiral's amplitude growth, however. 

A deeper analysis of the non-linear effects associated with the secular transfer of angular momentum is beyond the scope of the present paper. 
In any case, we have shown that the average amount of angular momentum and energy exchanged in a given time step, resultant from the simulations, agree very well with those expected from theory. Even though the intermediary processes that lead to the dissipation of orbital energy when a star interacts with the spiral wave are not reproduced by our simulations, it seems that the final torque applied by the spirals on the stellar orbits that contribute to the spiral response density is reasonably accounted for by our experiments. This enables us to go further in the study of the effects of the spiral perturbation on the stellar orbits, and specifically, the effects of the corotation resonance on the surface density of the stellar disk.      

\subsection{The minimum of stellar density at corotation}
\label{min_star_density}

Since we can consider the surface density of the stellar disk being made of the superposition of stellar orbits, a study of the changes of the orbital response to a perturbation is ultimately a study of the evolution of the disk surface density. The secular changes in the mean orbital radii are the main responsible for the formation of the minimum in the stellar density at the corotation radius, as we can see in Figs.~\ref{fig:density_t5Gyr} and~\ref{fig:distr_dL_amprel}d. We have also seen that the orbital changes of stars trapped in horseshoe orbits, during the linear growth of the perturbation, cause only small changes of the initial stellar density at corotation (cf. Fig.~\ref{fig:distr_dL_amprel}c). The minimum density at corotation, obtained with the spiral parameters of the model Sp1 ($f_{r0}=0.075$), reaches a relative amplitude $\approx 0.35$ at the time step $t=5$ Gyr, and is followed by a maximum of stellar density $\approx 0.18$ at a radius $r\sim 10.3$ kpc. We believe that such maximum is initially formed by some stars that got trapped in horseshoe orbits and that spend more time in the outer side of the orbit (relative to the corotation radius) than in the inner side of it. During the steady phase of the spirals, the amplitude of this maximum increases due to the increase of the mean orbital radii in regions outside corotation. We cannot also exclude the dependence of this maximum on the orbital changes due to the effects of the 4:1 OLR. We have also run other simulations with the same set of parameters but changing the value of $f_{r0}$: a simulation run with $f_{r0}=0.05$ produced a minimum at corotation with a quite small relative amplitude, $\approx 0.07$; otherwise, a simulation run with $f_{r0}=0.1$ generated a larger minimum with relative amplitude $\approx 0.5$, followed by a maximum of $\approx 0.25$. The configuration of maxima and minima shown in Fig.~\ref{fig:distr_dL_amprel}d reaches a quasi-steady state within a few time steps before the end of the simulation (at $t=5$ Gyr). We performed a simulation with the same initial conditions but for a total integration time of 10 Gyr and small further variations in the distribution of the relative amplitude of stellar density were observed after the step $t=5$ Gyr, the largest one being an increase of the minimum at corotation from 0.35 to $\sim 0.40$. We observed that the main variations in the stellar surface density occurred in a time interval of 3 Gyr, which is the interval when the spirals remain in a stationary state in our simulations. In this paper, we concentrate on the minimum associated with the corotation resonance and the maximum just beyond it; the features in the density that seem to be associated with the 4:1 OLR and the 2:1 OLR are briefly discussed in $\S$~\ref{add_remarks}.

We emphasize here that we are reporting a minimum in the density at corotation which is formed only in the stellar component of the disk. However, a minimum in the density of the gas component is also present. Indeed, a ring-like region deficient in neutral hydrogen, with radius slightly greater than the solar Galactic radius, can be seen 
in the results of H\,{\sevensize I} surveys of the past, eg.  \citet{Kerr1969},  \citet{Burton1976} (see fig. 6 of that paper). ALM mapped the ring and interpreted it as being an effect of the corotation resonance.
 L+8 explains that the void is caused by the `pumping out' effect of the spiral structure that produces the flow of the gas in opposite directions from corotation: an inward flow inside and an outward flow outside it. Both analytical and numerical studies of the gas flow near corotation have predicted such behaviour (e.g. \citealt{Lacey_Fall1985}; \citealt{Lepine2001}). A direct observation of this process was recently conducted by \citet{Elmegreen2009} from the analysis of the flow of gas in the corotation region of the galaxy NGC 1365.

Therefore, the minimum of mass density at corotation seem to be formed by both a depletion in the density of the gas and the stellar components. The existence and formation of the minimum in the density of gas had already been investigated by the studies above mentioned. Here, we propose that the minimum of density of the stellar component can be explained by the long-term evolution of the mean orbital radii of the stars. The underlying physics behind the formation of the minima in the density of gas and stars at corotation are different, mainly due to the different mean free path of these two components. However, the final results of these processes point to a common picture for the density of the disk at the corotation circle. 

\subsection{Results from simulations using different spiral models}
\label{other_spiral_models}

\begin{figure*}
	\includegraphics[scale=0.49]{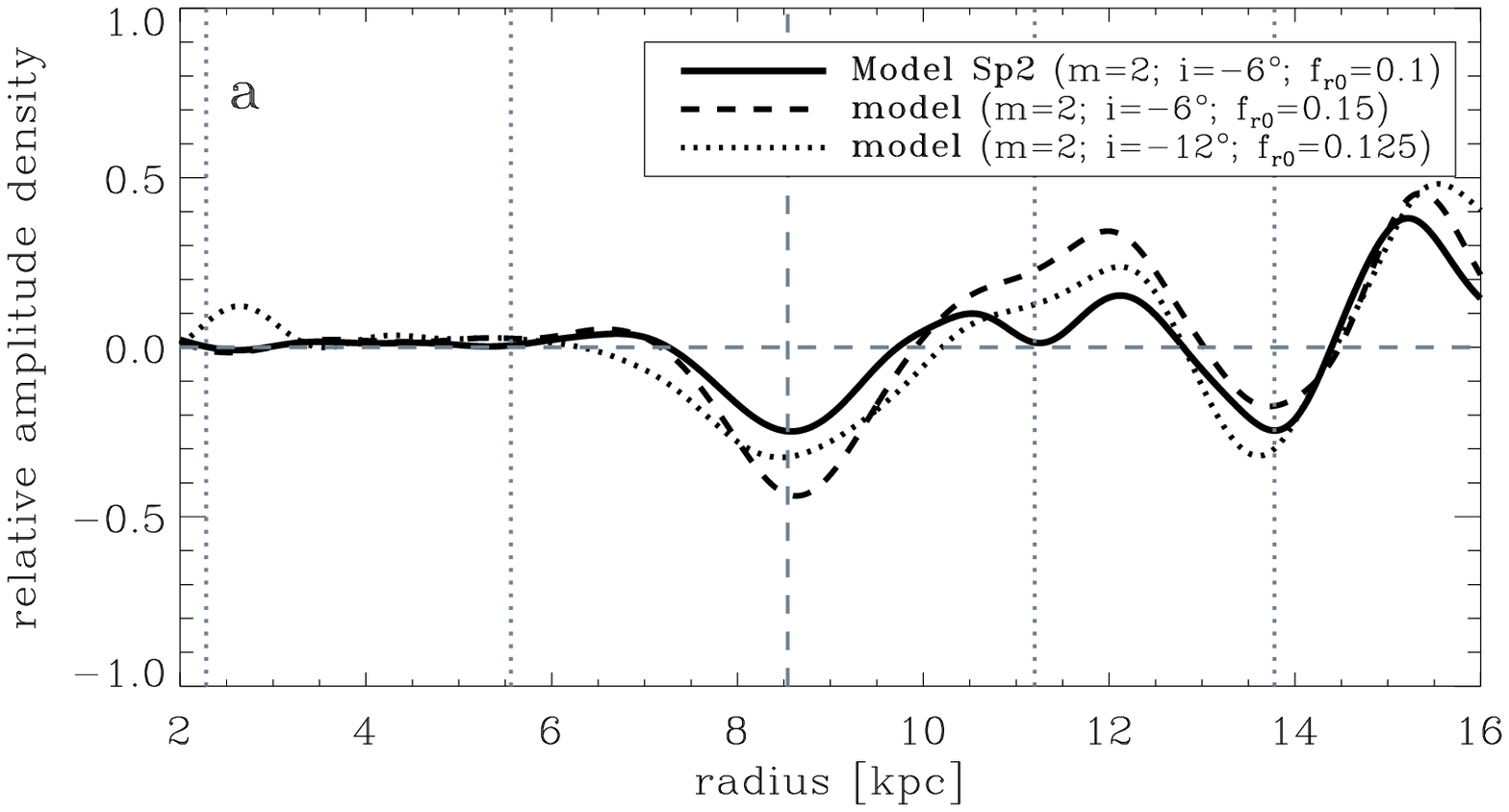}
	\includegraphics[scale=0.49]{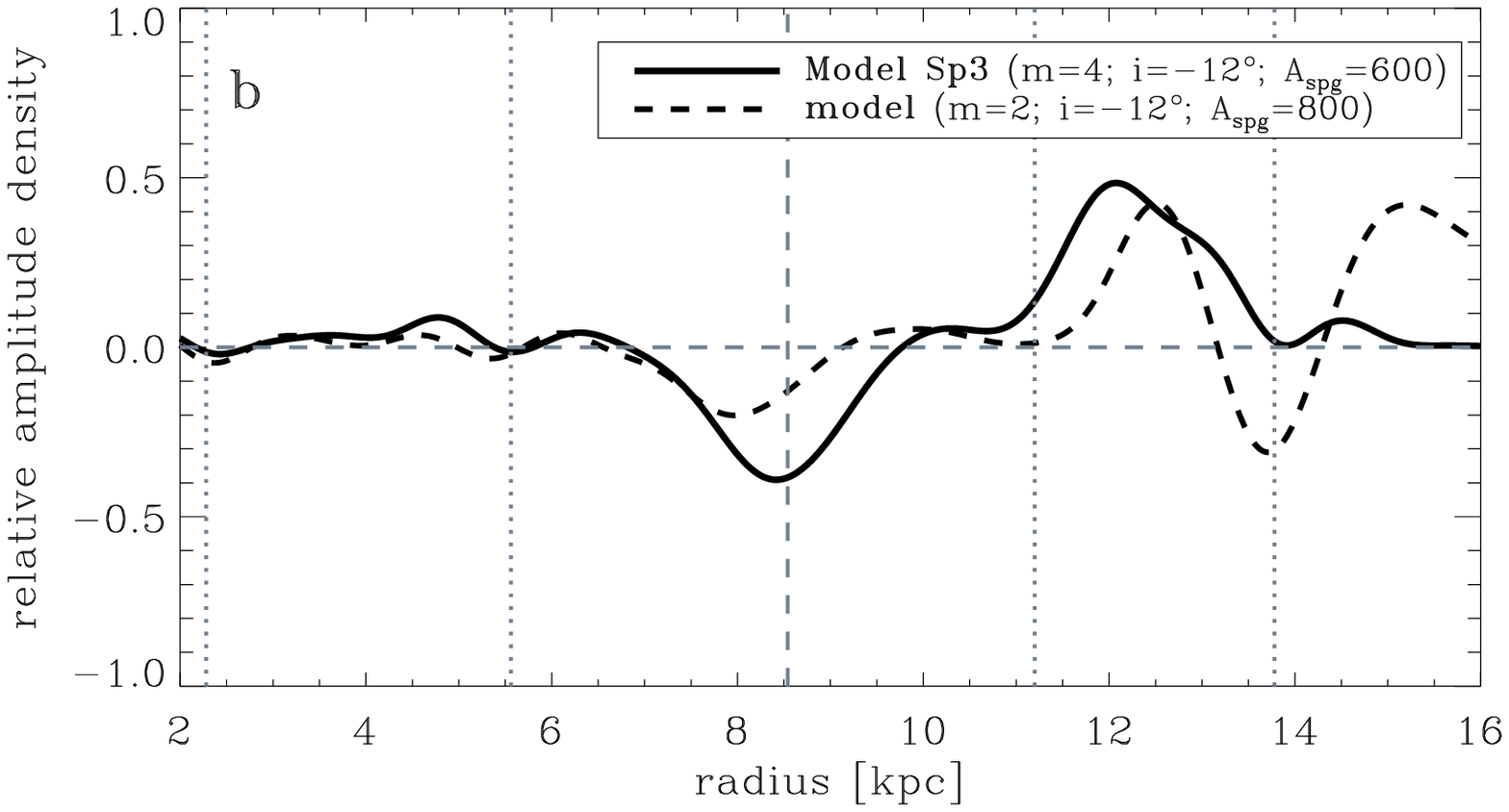}
	\includegraphics[scale=0.49]{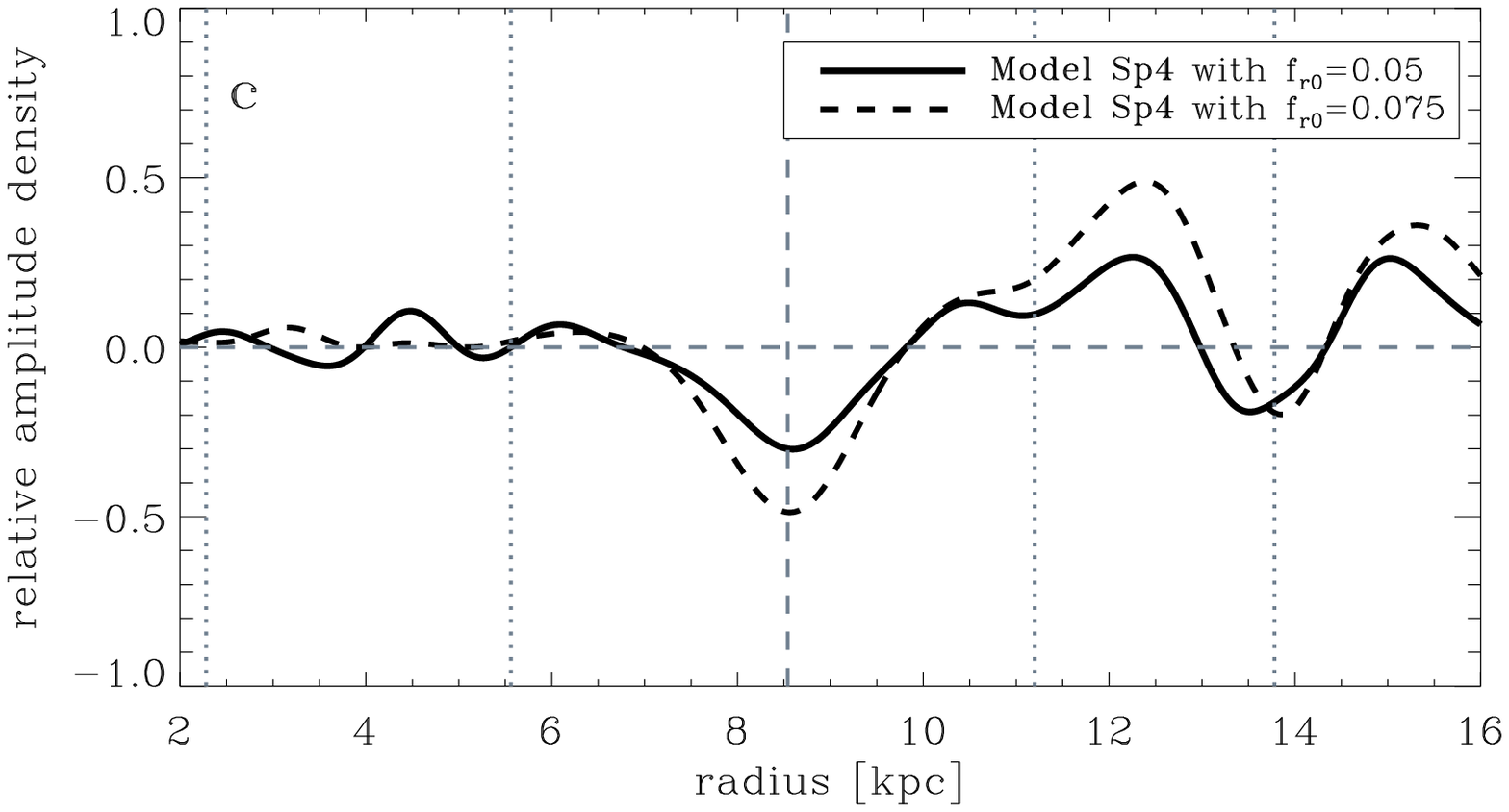}
	\includegraphics[scale=0.49]{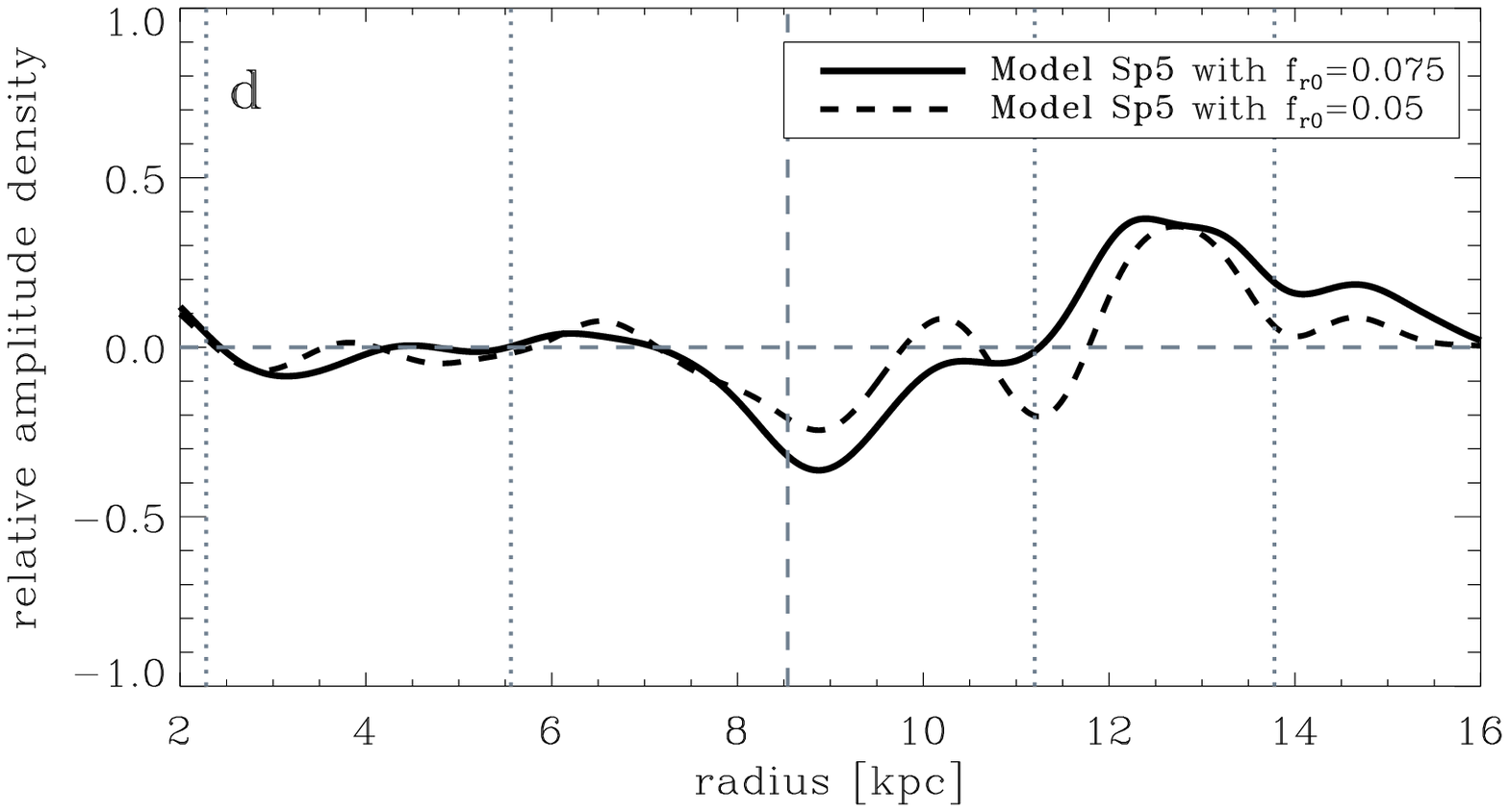}
	\includegraphics[scale=0.49]{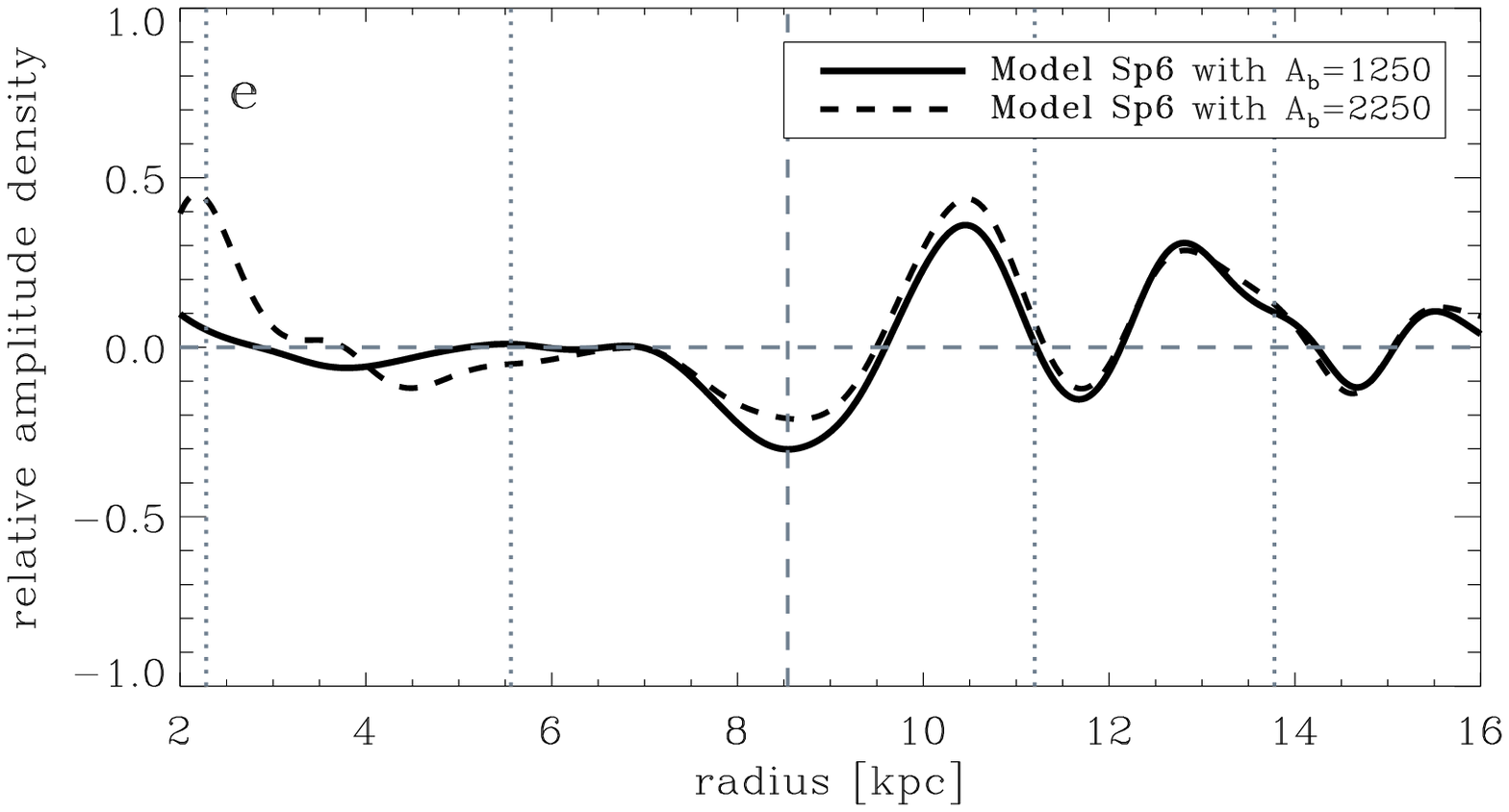}
	\includegraphics[scale=0.49]{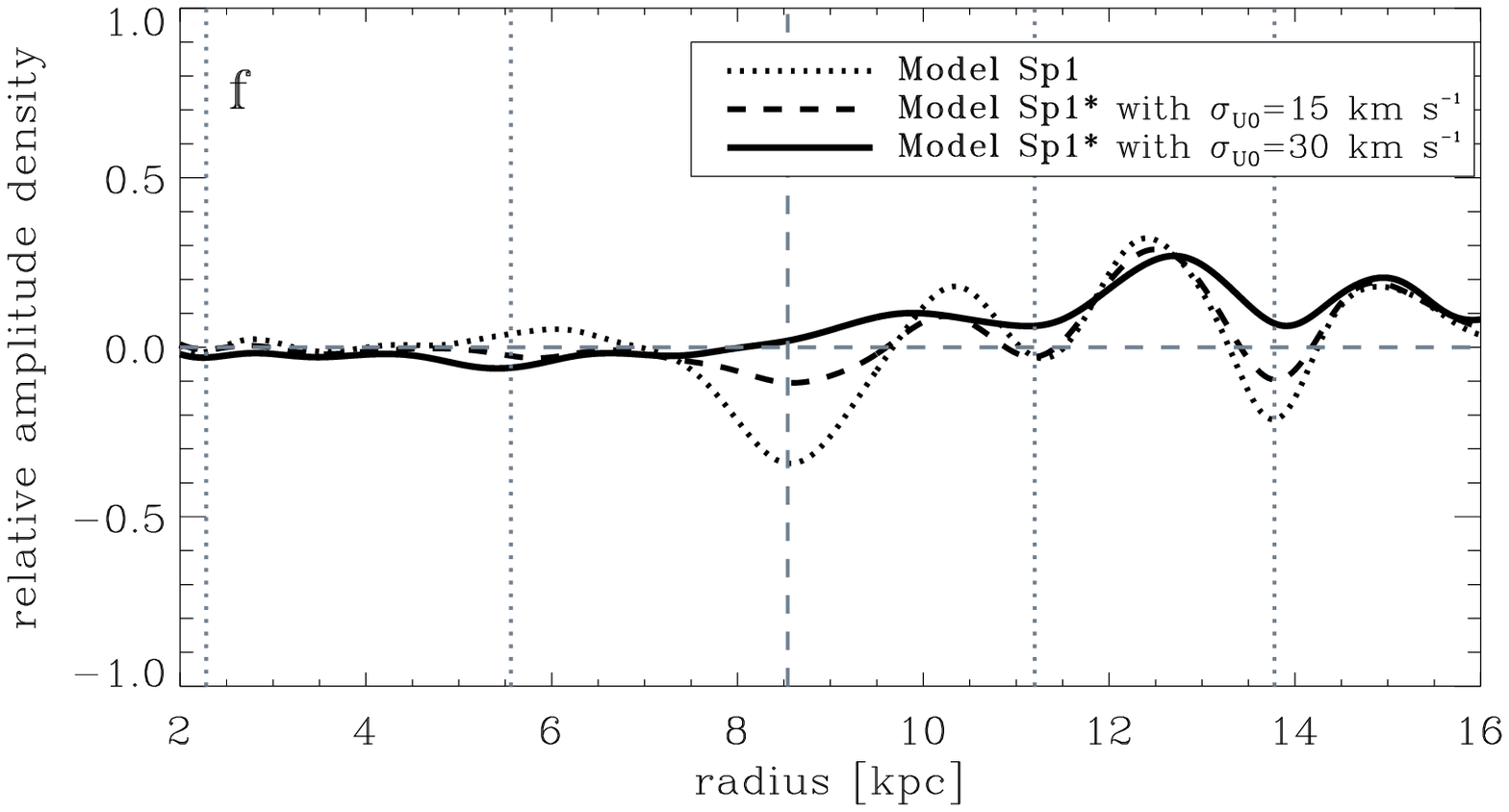}
	\caption{Radial distributions of the relative amplitude of density variations taken at the end of the simulations ($t=5$ Gyr). The panels a, b, c, d and e show the results of simulations using the spiral models Sp2, Sp3, Sp4, Sp5 and Sp6, respectively, from Table~\ref{tab:models_non-axis}. The variants of each model have their parameters labeled in the legend of each panel. The panel f shows the results of simulations using the spiral model Sp1, but for two cases of an initial hot stellar disk. The vertical lines indicate the positions of the resonances for a spiral pattern with angular speed $\Omega_{p}=25$ km s$^{-1}$ kpc$^{-1}$: corotation - dashed line; 2:1 ILR/OLR and 4:1 ILR/OLR - dotted lines. }
	\label{fig:KDE_amprel_distrib}
	\end{figure*}

We next present brief comments on the results about the formation of the minimum density at corotation and for the simulations using the spiral models from Table~\ref{tab:models_non-axis}, as well as for some variants of such models. We restrict here to the analysis of the minimum at corotation; a general discussion about the features in the density that occur at other radii is presented in $\S$~\ref{add_remarks}. Figure~\ref{fig:KDE_amprel_distrib} shows the radial distributions of the relative amplitude of density variations for simulation runs using the spiral models Sp2, Sp3, Sp4, Sp5 and Sp6 (shown as solid curves), in the same way as the one shown in Fig.~\ref{fig:distr_dL_amprel}d for the model Sp1. We also discuss a variation of the model Sp1 (Sp1$^{*}$) which simulates a stellar disk with hotter initial conditions. All the distributions were taken at the time step $t=5$ Gyr.

\begin{itemize}
\item {\bf Model Sp2.} Figure~\ref{fig:KDE_amprel_distrib}a shows the distributions obtained with a single 2-armed spiral mode: the solid curve shows the results using the model Sp2 - the relative amplitude of density variation at the minimum is $\approx 0.25$. The dashed curve represents the distribution for a model with the parameters ($m=2$, $i=-6^{\circ}$, $f_{r0}=0.15$) - the greater value for $f_{r0}$ leads to a larger amplitude of the minimum of $\approx 0.43$. The dotted curve shows the distribution for a model with ($m=2$, $i=-12^{\circ}$, $f_{r0}=0.125$); the relative amplitude of the minimum is $\approx 0.32$. Comparing these distributions to that from the model Sp1, we see that a 4-armed structure with a field strength $f_{r0}=0.075$ is able to produce a minimum in the stellar density with a larger amplitude than a 2-armed structure with $f_{r0}=0.1$.

\item {\bf Model Sp3.} This model employs the spiral potential with Gaussian profile formulated by JLBB. The model presents minor changes from the original values of the parameters given by the authors, which are specifically the number of arms $m=4$ and pitch angle $i=-12^{\circ}$. These changes are intended to make easier the comparison between the results from this model and from the model Sp1, which uses the cosine profile for the spiral potential. 
We have kept the arms half-width $\sigma_{g}$ constant along the radius, as in the original paper. The relative density variations using the model Sp3 are shown as the solid curve in Fig.~\ref{fig:KDE_amprel_distrib}b. The profile of the minimum at corotation is quite similar to the one obtained with the model Sp1; the relative amplitude of the minimum is 0.40. As argued by JLBB, this new model of the spiral potential leads to a more realistic description of the Galactic spiral structure in several aspects, and our findings here that it is also capable to produce a minimum in the stellar density at corotation corroborate its importance to the study of the stellar dynamics influenced by the spiral perturbation. The dashed curve in Fig.~\ref{fig:KDE_amprel_distrib}b is the result of a simulation that also employs Gaussian spiral arms, but with $m=2$ arms, pitch angle $i=-12^{\circ}$, and a larger spiral amplitude $\mathcal{A}_{\mathrm{sp_{g}}}=800$ km$^{2}$ s$^{-2}$ kpc$^{-1}$, which implies in a ratio $f_{r0}=0.06$ according to JLBB. Despite the relatively smaller amplitude, the minimum density is displaced from corotation, occurring at a smaller radius of $\approx 7.9$ kpc.

\item {\bf Model Sp4.} This model is based on the self-sustained model of the Galactic spiral structure proposed by \citet{Lepine2001}, which consists of a superposition of 2- and 4-armed modes with pitch angles $-6^{\circ}$ and $-12^{\circ}$, respectively. We adopt the same pattern speed $\Omega_{p}=25$ km s$^{-1}$ kpc$^{-1}$ for both of the modes. Adopting the same ratio of field strength $f_{r0}$, the amplitudes of both modes present the same value at each radii. The solid curve in Fig.~\ref{fig:KDE_amprel_distrib}c shows the relative density variations using the configurations of the model Sp4 and for a ratio $f_{r0}=0.05$; the amplitude of the minimum at corotation reaches 30\% of the underlying density. The dashed curve shows the variations for a ratio $f_{r0}=0.075$, with the amplitude of the minimum reaching 50\% in this case. The latter distribution can be compared to the results using the model Sp1, shown in Fig.~\ref{fig:distr_dL_amprel}d, where the same configuration for the single 4-armed mode produced a minimum at corotation with a relative lower amplitude of 35\%.

\item {\bf Model Sp5.} Some recent studies relate the positions of the Hyades and Sirius moving groups in the {\it U-V\/} velocity plane of the solar neighbourhood to orbital families induced by the 4:1 ILR of a 2-armed spiral structure (e.g. \citealt{Quillen_Minchev2005,Famaey2007,Pompeia2011}). A single 2-armed mode, with its 4:1 ILR close to the solar orbit radius, would imply in a corotation radius at $\sim 11-12$ kpc (a model with these properties was recently proposed by \citealt{Siebert2012} from the analysis of the Galactocentric radial velocity field of stars from the {\it RAVE\/} survey). Examining the distributions in Fig.~\ref{fig:KDE_amprel_distrib}a, no strong feature in the density is evidenced near the 4:1 ILR of a pure two-armed mode (at $r=5.6$ kpc), even with the largest value of $f_{r0}$ employed in the simulations. Therefore, we do not expect that a single spiral mode $m=2$, with the 4:1 ILR near the solar radius and corotation at $\sim 11-12$ kpc, could account for the minimum in the disk surface density, with radius between 8.0 and 8.5 kpc, that is likely the cause of the dip in the rotation curve at this region. To reconcile these two scenarios, L+8 has raised the hypothesis of a possible existence of multiple spiral patterns with different pattern speeds: while the main grand-design spiral pattern has its corotation at $8.3-8.4$ kpc, it could coexist with a slower $m=2$ pattern whose 4:1 ILR smoothly connects with the corotation of the main pattern. This has motivated us to investigate the effects of such superposition of two spiral modes on the disk density: a $m=4$ pattern with $\Omega_{p}=25$ km s$^{-1}$ kpc$^{-1}$, and a $m=2$ pattern with $\Omega_{p}=18$ km s$^{-1}$ kpc$^{-1}$. We adopt the same pitch angle $i=-12^{\circ}$ for both modes, which results in a 2-armed pattern with amplitude two times greater than the 4-armed one. The distributions of density variations using the model Sp5 are shown in Fig.~\ref{fig:KDE_amprel_distrib}d, for $f_{r0}=0.05$ and 0.075 (dashed and solid curves, respectively). The relative amplitudes of the minimum density at corotation are similar to those found with the other models, but the minimum is displaced from corotation to a slightly larger radius of $\approx 8.8$ kpc.

\item {\bf Model Sp6.} Here we investigate the influence of the perturbation from the central bar when it acts together with that from the spiral arms of the model Sp1. For the gravitational potential due to the bar, we use the analytical form given by \citet{Dehnen2000}:

\begin{equation}
\label{eq:potbar}
\Phi_{b}(r,\theta';t)=A_{b}\cos [2(\theta'-\Omega_{b}t)]\times \left\{
\begin{array}{ll}
\left(\frac{r}{r_{b}}\right)^{3}-2, & \, r\leq r_{b}\\
-\left(\frac{r_{b}}{r}\right)^{3}, & \, r\geq r_{b},
\end{array}
\right.
\end{equation}
where $A_{b}$ is the amplitude of the bar perturbation, which is dominated by the quadrupole term at $r\geq r_{b}$; $r_{b}$ is the bar half-length; and $\Omega_{b}$ is the bar pattern speed. We adopt a bar half-length $r_{b}=3.5$ kpc (e.g. \citealt{Lopez-Corredoira2001}; \citealt{Cabrera-Lavers2008}). We derive the amplitude of the bar perturbation through the parameter defined by \citet{Dehnen2000} as $\alpha=3(A_{b}/V^{2}_{0})(r_{b}/R_{0})^{3}$, which is the ratio of the forces due to the bar's quadrupole and the axisymmetric background at Galactocentric radius $R_{0}$ on the bar's major axis. We perform tests with $A_{b}=1250$ and 2250 km$^{2}$ s$^{-2}$, which correspond to the values $\alpha=0.007$ and 0.013 used by \citet{Dehnen2000}. We also investigate the influences of a stronger bar, with amplitude $A_{b}=4500$ km$^{2}$ s$^{-2}$. This value is consistent with the parameter $|\epsilon_{b}|=0.03$ (the same as $\alpha$) used by MF10 and based on the results from \citet{Rodriguez-Fernandez_Combes2008}. 

Different methods have led to different values measured for the pattern speed of the bar (\citealt[and references therein]{Gerhard2011}), ranging from $\Omega_{b}=30$ to 60 km s$^{-1}$ kpc$^{-1}$.
Here we investigate the case for $\Omega_{b}=40$ km s$^{-1}$ kpc$^{-1}$, not because we believe this is the most probable value, but only because the 4:1 OLR of the bar takes place near the corotation radius of the spirals; for the same reason, we also ran simulations with $\Omega_{b}=50$ km s$^{-1}$ kpc$^{-1}$, which puts the bar's 2:1 OLR close to the spiral's corotation. These values enable the study of the effects of resonance overlap (e.g. MF10) and its influence on the minimum density formed at the corotation of the spiral pattern.

Preliminary tests with the bar perturbation have led us to opt for minor changes in the initial conditions of the disk and the axisymmetric potential: since the rotation curve from Eq.~\ref{eq:v_L2011} may already include the contribution of the bar potential in the central regions of the Galaxy, we have retrieved only the part of the curve relative to the disk (the first term on the right-hand side of Eq.~\ref{eq:v_L2011}) for the derivation of the axisymmetric component of the potential, but slightly changing the values of the parameters to keep the same values for $V_{0}$ and $R_{cr}$ as in the other simulations; we also increased the initial velocity dispersion in the central region of the disk, but keeping the value $\sigma_{U_{0}}=5$ km s$^{-1}$ at the solar orbit radius. These changes are intended to decrease the initial strong stellar response to the bar perturbation, which would lead to intense modifications of the stellar orbits in the central regions. The distributions of density variations using the model Sp6 are shown in Fig.~\ref{fig:KDE_amprel_distrib}e. The solid line shows the distribution using the bar amplitude $A_{b}=1250$ km$^{2}$ s$^{-2}$ and the pattern speed $\Omega_{b}=40$ km s$^{-1}$ kpc$^{-1}$; these values do not alter significantly the structure of the minimum density at the corotation of the spiral pattern, only slightly decreasing its relative amplitude to $\sim 0.30$. However, with the larger bar amplitude $A_{b}=2250$ km$^{2}$ s$^{-2}$ and same $\Omega_{b}$ (dashed line in Fig.~\ref{fig:KDE_amprel_distrib}e), the relative amplitude of the minimum density at the spiral's corotation decreases further to $\sim 0.20$. The trend of decrease of the amplitude of the minimum density with the increase of the amplitude of the bar perturbation was also verified with a simulation run using $A_{b}=4500$ km$^{2}$ s$^{-2}$; in this case almost no variations in the stellar density were observed near the corotation radius of the spiral pattern (not shown in Fig.~\ref{fig:KDE_amprel_distrib}e). However, in this last case, a strong depopulation of stellar orbits in the central regions of the disk was observed, as well as an intense modification of the initial exponential profile of the disk surface density at the end of the simulation. This may be the result of strong stellar radial migration promoted by the overlap between the corotation of the bar and the 4:1 ILR of the spiral pattern (at $r\sim 5.5$ kpc), as well as between the 2:1 ILRs of both the bar and spiral patterns (at $r\sim 2.2$ kpc) (MF10). Similar results were obtained from simulations using the smaller values for $A_{b}$, but with the faster bar pattern speed $\Omega_{b}=50$ km s$^{-1}$ kpc$^{-1}$. We regard these aforementioned last results as unphysical, since it is most likely that our simple simulations are not appropriate to reproduce the response of the particles to the interactions involving strong and fast non-axisymmetric structures. 

\item {\bf Model Sp1$^{*}$.} This model employs the same spiral configuration of the model Sp1, with the changes made only in the initial distribution of velocity dispersion of the disk stars, in the sense to simulate initially hotter disks. The dotted line in Fig.~\ref{fig:KDE_amprel_distrib}f shows the relative density variations using the spiral model Sp1 (already shown in Fig.~\ref{fig:distr_dL_amprel}d), which was the result of a disk with an initial radial velocity dispersion profile normalized to $\sigma_{U_{0}}=5$ km s$^{-1}$ at $R_{0}$ (dash-dotted curve in Fig.~\ref{fig:sigVr_simulat}). The dashed line in Fig.~\ref{fig:KDE_amprel_distrib}f represents the density variations for a disk with a similar initial radial velocity dispersion profile, but normalized to $\sigma_{U_{0}}=15$ km s$^{-1}$. The curve shows that the amplitudes of density variations are smaller when compared to the previous one, mainly with respect to the minimum at corotation, which in this case reaches only $\sim 10\%$ of the underlying density. The minimum density at corotation disappears drastically in a simulation with the initial radial velocity dispersion normalized to $\sigma_{U_{0}}=30$ km s$^{-1}$. This latter case is shown by the distribution represented by the solid line in Fig.~\ref{fig:KDE_amprel_distrib}f. These trend shows that the minimum density at corotation is the result of an evolutive process that strictly acts on an initially cold stellar disk. These are not unexpected results, since the increase in the velocity dispersion leads to a smaller response to the spiral perturbation. 
The secular stellar heating, as well as the secular increase in the Toomre's $Q$-parameter of the stellar disk, must lead to a saturation in the process of angular momentum and energy transfer as the system evolves. As discussed in $\S$~\ref{rad_distrib}, even though it is not apparent from Eq.~\ref{eq:drmed_dt}, the evolution of the properties of the disk basic state (surface density, velocity dispersion, epicyclic frequency), as well as the properties of the spiral pattern (amplitude, pitch angle) must dictate the saturation process in the exchange of $L$ and energy $E$ in a self-consistent scenario. As shown by Z98, the spiral modes usually present in the advanced stages of secular evolution have small amplitudes and pitch angle, which decreases the rate of evolution and leads to patterns that could last for a timescale on the order of a Hubble time.

\end{itemize}

\subsection{The disk heating at the solar orbit radius}
\label{disk_heating}

	\begin{figure}
	\includegraphics[scale=0.48]{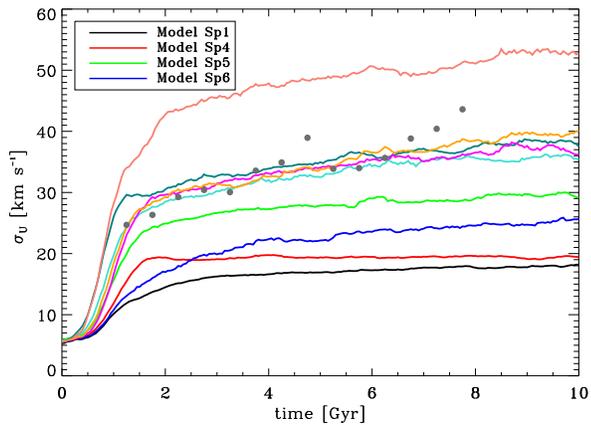}
	\caption{Time evolution of the radial velocity dispersion at the solar orbit radius $R_{0}$ from simulation runs with the models Sp1, Sp4, Sp5, Sp6 and for other models described in $\S$~\ref{disk_heating} (see text). The dots are the observed age-velocity dispersion relation of the solar neighbourhood stars derived from the Geneva-Copenhagen Survey data (\citealt{Holmberg2009}).}
	\label{fig:disk_heating_R0}
	\end{figure}

One point linked to the previous discussion about the heating of the stellar disk is that, as one can see from Fig.~\ref{fig:sigVr_simulat}, the simulation with the spiral model Sp1 produced a radial velocity dispersion at the solar radius $R_{0}$ with a value too low, $\sim 15$ km s$^{-1}$. Observations of the solar neighbourhood stars show a velocity dispersion $\sigma_{U_{0}}\sim 35$ km s$^{-1}$ for a population with mean age $\sim 5$ Gyr (\citealt{Holmberg2009}). On the other hand, we have shown that the model Sp1$^{*}$, with initially hot disks, is more stable against the spiral perturbations, producing smaller amplitudes for the minimum density at corotation. 
Moreover, such minimum stellar density, if indeed exists, must be a global characteristic of the main old stellar disk population, not just of the coldest one. Thus we face the problem of how to conciliate the observational evidences for: 1 - a minimum density at corotation which is indirectly verified in the dip of the rotation curve and must affect globally the main old stellar population; and 2 - an old stellar population which mostly presents high velocity dispersions in the solar vicinity. It has also to be noted that the secular heating that we detect in our simulations with the model Sp1 ($\S$~\ref{reliable_simulat}) does not seem to be able to account for the observed velocity dispersion. This could be in part due to the fact that we used the current epoch observed values of the spiral parameters, which were kept fixed over the entire simulation, with the exception of the amplitude of perturbation. Z99 was able to reproduce the age-velocity dispersion of the solar neighbourhood using average values for the parameters of an expression that gives the amount of secular heating with time (see eq. 9 from the referred paper). However, in that case, the author used a pitch angle of $i=20^{\circ}$ ($12^{\circ}$ in our case), and a factor $\Omega(R_{0})-\Omega_{p}=8$ km s$^{-1}$ kpc$^{-1}$, which in our models is only about 3.6 km s$^{-1}$ kpc$^{-1}$.  

Some scattering mechanisms have been proposed in the literature to explain the relation between the age and the velocity dispersion of stars in the solar neighbourhood, with the agent of scattering being, e.g., transient spiral waves, giant molecular clouds, halo black holes, infall of satellite galaxies, among others. 
\citet[hereafter MQ06]{Minchev_Quillen2006} proposed a new mechanism of stellar heating by the resonance overlap of two steady-state spiral waves moving at different pattern speeds; MF10 studied the case of stellar heating due to bar-spiral interactions. In this section we investigate in what situations a long-lived spiral pattern is able to promote not only the formation of the minimum density at corotation but also the observed stellar heating at $R_{0}$. 

Fig.~\ref{fig:disk_heating_R0} shows the time evolution of the radial velocity dispersion $\sigma_{U}$ at the solar orbit radius from simulation runs with several spiral models described in the previous section, as well as for some of their variants which will be described below. All the curves of $\sigma_{U}(t)$ were taken by computing the square root of the bias-corrected variance of the radial velocity distribution of particles in an annulus centered at $R_{0}$ and with width $\Delta r=0.1$ kpc (MQ06); the computation was done for each time interval of 50 Myr ($\sim 1/4$ of the rotational period at $R_{0}$). The simulations were run for a total integration time of 10 Gyr. In order to better compare the results to the observational data, a little change was done in the expression for the growth rate of the amplitude of perturbation (Eq.~\ref{eq:amp_spiral_growth}), making the spirals grow from $t=0$ to $t=1.5$ Gyr. 
The four lowermost curves are (bottom to top): black for the model Sp1; red for the model Sp4 with $f_{r0}=0.075$; blue for the model Sp6 with bar amplitude $A_{b}=2250$ km$^{2}$ s$^{-2}$; light green for the model Sp5 with $f_{r0}=0.075$. The gray dots are the observed radial velocity dispersion of the solar neighbourhood stars derived from the Geneva-Copenhagen Survey data (\citealt{Holmberg2009}) and averaged over bins of 0.5 Gyr of the stellar age distribution (in the range $1-8$ Gyr); as done by \citet{Holmberg2009}, we use a subsample with age dispersion $\sigma_{\mathrm{Age}}<25\%$. 
All the four aforementioned curves give values of stellar heating rate well below the observed ones in the solar vicinity. However, it is clear that the models with multiple patterns moving at different speeds (Sp5 and Sp6) produced higher velocity dispersions when compared to the model with single pattern (Sp1) or two patterns with the same $\Omega_{p}$ (Sp4). These results are in agreement with those from MQ06. We then investigated some variants of the models Sp5 and Sp6 which resulted in larger heating rates, even for an initially cold stellar disk. Following the curves in Fig.~\ref{fig:disk_heating_R0}, the light blue curve shows the result of a model similar to the Sp5, but for stronger spiral perturbations, with $f_{r0}=0.1$ for both 2- and 4-armed modes. The magenta curve is for a model similar to the Sp6 but for the stronger bar, with $A_{b}=4500$ km$^{2}$ s$^{-2}$ (and $\Omega_{b}=40$ km s$^{-1}$ kpc$^{-1}$). The dark green curve is for a model also similar to the Sp5, also with $f_{r0}=0.1$, but exchanging the pattern speeds of the 2- and 4-armed modes: $\Omega_{p}=25$ km s$^{-1}$ kpc$^{-1}$ for the pattern with $m=2$, and $\Omega_{p}=18$ km s$^{-1}$ kpc$^{-1}$ for the pattern with $m=4$. This configuration now puts the 4:1 ILR of the 4-armed mode (which is a first-order resonance of this mode) close to the corotation of the 2-armed mode (which now is the main pattern with its corotation near $R_{0}$), and according to the figures 3 and 4 of MQ06, it also produces an extra amount of heating when compared to the configuration of the model Sp5. The orange curve is for a model also similar to the Sp6, but for a faster bar with $\Omega_{b}=50$ km s$^{-1}$ kpc$^{-1}$ (and $A_{b}=2250$ km$^{2}$ s$^{-2}$). These last four models seem to produce reasonable heating rates when compared to the observations, at least for the data points between $\sim 2$ and 6 Gyr. The heating rate beyond $\sim 6$ Gyr shown by the observed data is somewhat above the ones produced by the simulations. Despite these relative agreements, we have shown in $\S$~\ref{other_spiral_models} that the models with faster and stronger bars gave unreliable results for the disk surface density at the end of the simulations. Finally, the uppermost (salmon) curve in Fig.~\ref{fig:disk_heating_R0} is also for a bar/spiral model, but for both the stronger ($A_{b}=4500$ km$^{2}$ s$^{-2}$) and faster ($\Omega_{b}=50$ km s$^{-1}$ kpc$^{-1}$) bar case. This last model produced an exaggeratedly larger heating rate compared to the previous ones and to the actually observed, a result also found by MF10 with values for the bar and spiral parameters similar to the ones used here. From all the above results, it seems that the presence of two spiral patterns moving at different speeds is the best configuration to promote both the observed stellar heating at $R_{0}$ and the expected minimum stellar density at the corotation of the main pattern near $R_{0}$. According to MQ06, the extra heating achieved from such configuration is explained by the stochastic stellar motion due to the resonance overlap of the patterns. Here we suggest that, additionally to this mechanism, it is also possible that there has been an increase in the efficiency of secular heating in the studied case of two spirals with different $\Omega_{p}$, since the corotation of the slower pattern takes place farther from $R_{0}$, and according to Z99 it is expected that the function $\sigma_{U_{0}}(t)$ increases with the factor $\Omega(R_{0})-\Omega_{p}$.



\section[]{Observational evidences for the stellar density minimum}
\label{observ_evid}

\subsection{Indirect evidence from the minimum in the Galactic rotation curve}
\label{indirect_evid_rot_curve}

\begin{figure}
	\includegraphics[scale=0.50]{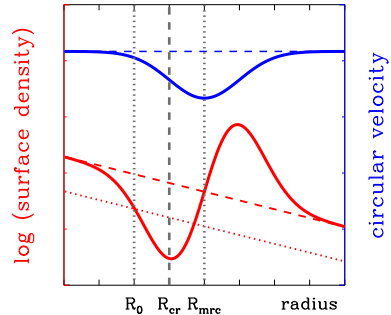}
	\caption{Schematic illustration of the radial profile of the disk surface density (red curve) and the correspondent circular velocity (blue curve) in the region near the corotation circle. Both profiles are given in arbitrary units. The vertical dashed line indicates the corotation radius $R_{\mathrm{cr}}$ and the vertical dotted lines indicate the solar Galactic radius $R_{0}$ and the radius of the minimum of the rotation curve $R_{\mathrm{mrc}}$.}
	\label{fig:surfdens_rotcurv}
	\end{figure}

A galactic disk with a ring of minimum density has implications for the total potential of its host galaxy. As commented by \citet{Sikora2012}, the disk surface density $\Sigma(r)$ and the centrifugal acceleration on a circular orbit $V^{2}_{c}/r$ in the disk plane are integral transforms of each other. Therefore, if a gap in the surface density distribution effectively exists at a given radius in the disk, a minimum in the rotation curve is expected slightly beyond it. Let us think, for instance, of an ideal gap with sharp edges, with zero density inside it. At the radius of the inner part of the gap, the rotation curve will start to decrease, and will reach a minimum at the radius of the outer edge. In this case, the minimum of the rotation curve is shifted from the radius of minimum density (the center of the gap) by half gap width. For the density profile shown in Figs.~\ref{fig:density_t5Gyr} and~\ref{fig:distr_dL_amprel}d, and considering only the minimum at corotation and the maximum just beyond it as the main changes from the initial distribution, we can fit them by the function:

\begin{equation}
\Sigma_{f}(r)=\Sigma_{i}(r)\left[1+\mathcal{C}\,\frac{\ud f_{\mathrm{mrc}}}{\ud r}\right],
\label{eq:Sigma_function}
\end{equation}
where we denote $\Sigma_{i}$ and $\Sigma_{f}$ as the initial and final surface densities, respectively; $\mathcal{C}$ is an arbitrary constant necessary to the conversion of units and to the adjustment of the amplitude of the minimum density; $f_{\mathrm{mrc}}$ is the Gaussian function expressed in Eq.~\ref{eq:funct_mrc}. Figure~\ref{fig:surfdens_rotcurv} shows the plot for $\Sigma_{f}(r)$ as the red curve and in arbitrary units. The circular velocity resultant from a disk with the surface density profile of $\Sigma_{f}$ is shown as the blue curve in Fig.~\ref{fig:surfdens_rotcurv}, also in arbitrary units. As a consequence, the profile of the circular velocity can be fitted by the function $f_{\mathrm{mrc}}$. From Fig.~\ref{fig:surfdens_rotcurv}, we see that the radius of the minimum in the rotation curve $R_{\mathrm{mrc}}$ (indicated by the vertical dotted line on the right) is shifted from the corotation radius (the radius of minimum density - vertical dashed line) by the half-width $\sigma_{\mathrm{mrc}}=R_{\mathrm{mrc}}-R_{cr}$. In the above example, the same amplitude for the minimum and the maximum in the profile of the density is considered. In the case of the profile resultant from the simulation, where the amplitude of the maximum is smaller than that of the minimum, we could fit it using a Gaussian with a slight positive skewness for the function $f_{\mathrm{mrc}}$.

\citet*[hereafter SHO]{Sofue2009} presented a unified rotation curve of the Galaxy and its decomposition into three main mass components. According to the authors, the so-called {\it 9-kpc dip is exhibited as the most pronounced and peculiar feature in the Galactic rotation curve\/}. This local minimum in the rotation curve is observed in each plot of many different tracers of the rotation pattern of the Galaxy. The outer rotation curves based on H\,{\sevensize II}/CO data (e.g. \citealt{Georgelin_Georgelin1976}; \citealt{Clemens1985}, \citealt{Fich_Blitz_Stark1989}), maser observations from high-mass star-forming regions (\citealt{Honma2012}), as also C-stars (\citealt{Demers_Battinelli2007}), reveal the feature at $\sim9$ kpc in all of the cases. A rotation curve based on planetary nebulae and AGB stars derived by \citet{Amaral1996} also exhibits a minimum at $\sim 8.5$ kpc. In addition, \citet{Lepine2008} verified that a rotation curve with a minimum at $\sim 8.8$ kpc well explains the radial distribution of the observed epicyclic frequency $\kappa$ determined from the initial velocity distribution of open clusters. Because of the different nature of these tracers (stars or atomic and neutral gas), one can conclude that the minimum in the rotation curve can neither be solely due to non-linear response of the gas to the spiral perturbation, nor simply local deviations of stars from the circular motion (SHO). In addition, this feature is also observed in the rotation curve obtained by \citet{Honma_Sofue1997} from the H\,{\sevensize I} thickness method of Merrifield (\citealt{Merrifield1992}). This method is applied to the entire H\,{\sevensize I} disk, avoiding the bias effect pointed out by \citet{Binney_Dehnen1997} that is introduced when the way the rotation velocity is measured changes from one to the other side of the solar circle.

In order to reproduce such dip in the rotation curve, SHO superposed a massive wavy ring on the exponential density profile of the disk. The ring has a radial profile similar to the one shown in Figure~\ref{fig:surfdens_rotcurv} for the surface density: it consists of a minimum at 8.5 kpc, a maximum at 11 kpc, a wave node at 9.5 kpc, and an amplitude $\sim 0.34$ times the underlying surface density for both minimum and maximum (the authors used $R_{0}=8$ kpc). Perhaps, this is the first proposal of such a structure in the Galactic disk. We see a great resemblance between this structure and the ringed pattern of the stellar density variations obtained from our simulations. In our case, the minimum is located exactly at the corotation radius (8.54 kpc), followed by a maximum with lower amplitude at $r\sim 10.3$ kpc. The amplitude of the minimum of $\sim 35\%$ of the underlying density, resultant from the simulation using the spiral model Sp1, closely agrees with the one proposed by SHO. However, since our simulations deal only with the stellar component, we also have to consider the contribution of the minimum in the gas density. For this, we performed a rough estimation based on the density contrast of the gap of H\,{\sevensize I} given by ALM; the authors derived a depletion at the gap of $\sim 80\%$ of the mean H\,{\sevensize I} density of regions close to the gap. For the ratio between the surface density of H\,{\sevensize I} $\Sigma_{\mathrm{H{\scriptscriptstyle I}}}$ and the surface density of stars $\Sigma_{\mathrm{\scriptstyle star}}$ at corotation, we used the value $\Sigma_{\mathrm{H{\scriptscriptstyle I}}}/\Sigma_{\mathrm{\scriptstyle star}}=0.23$, which is close to the ratio obtained from the model of \citet{Flynn2006} for the local mass surface density. We concluded that a relative variation of only $\sim 20\%$ of the total disk density would be produced considering only the minimum in the density of the gas at corotation. To produce a relative variation in the range $\sim 30\% - 40\%$ of the total density, we still need to employ a relative variation of $\sim 20\% - 30\%$ in the density of the stellar component. We have to note that the dispersion of the observational data around the dip in the rotation curve could even support a disk with a minimum in the total density of the order of 50\% of the underlying density.    


\subsection{Direct evidence from the Galactic distribution of old stellar objects}
\label{direct_evid_rot_curve}

	\begin{figure*}
	\includegraphics[scale=0.42]{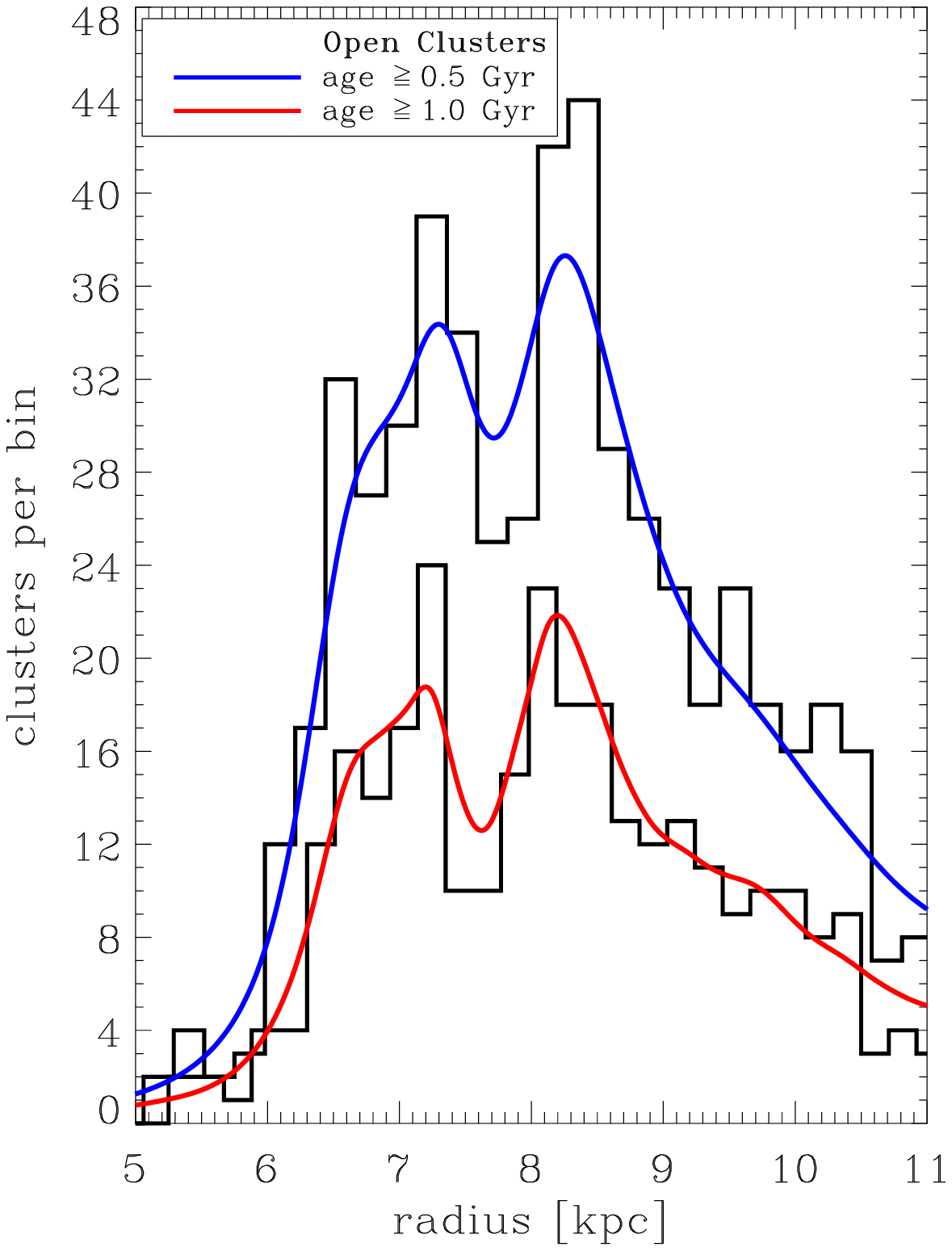}
	\includegraphics[scale=0.42]{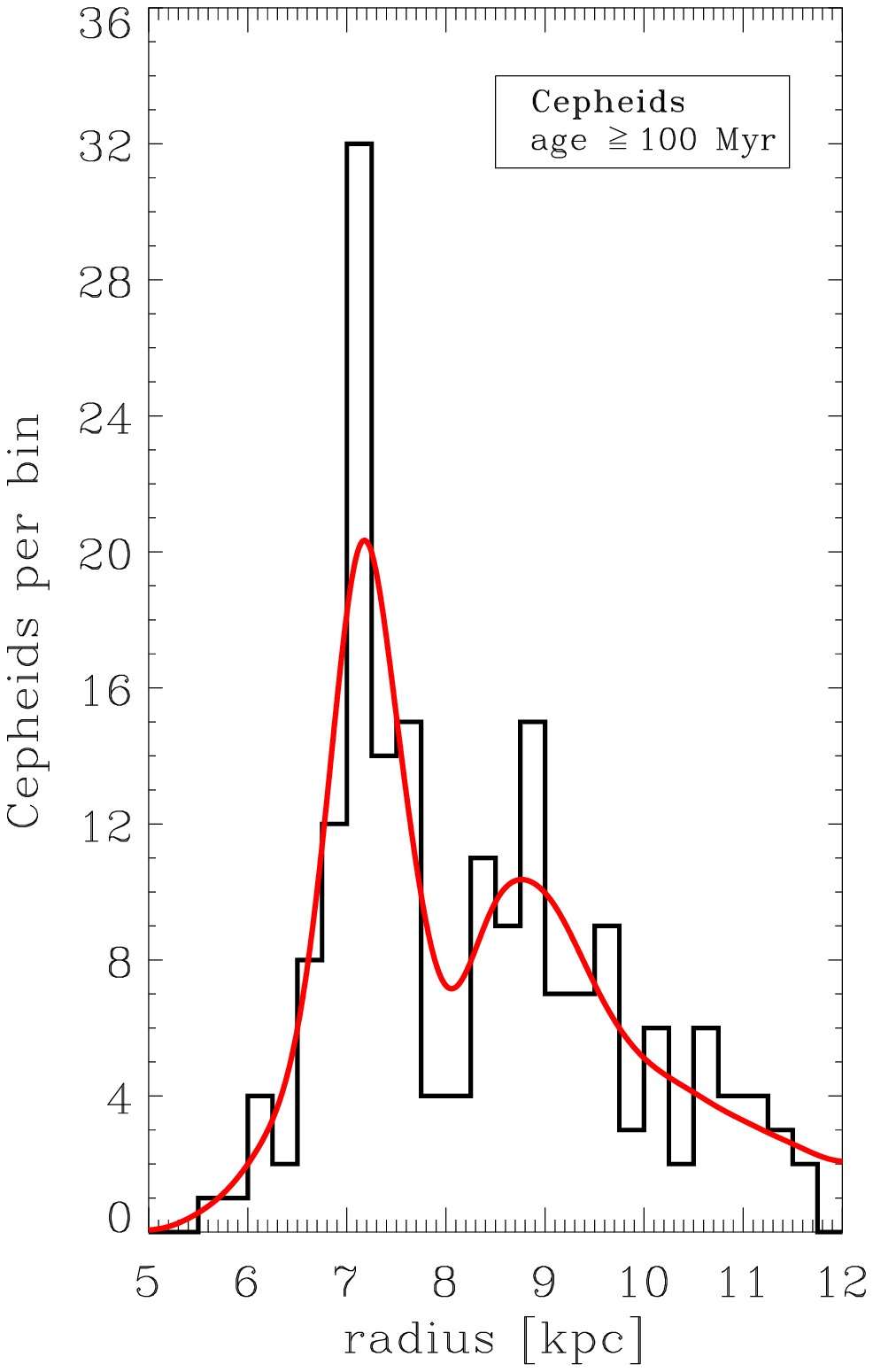}
	\includegraphics[scale=0.42]{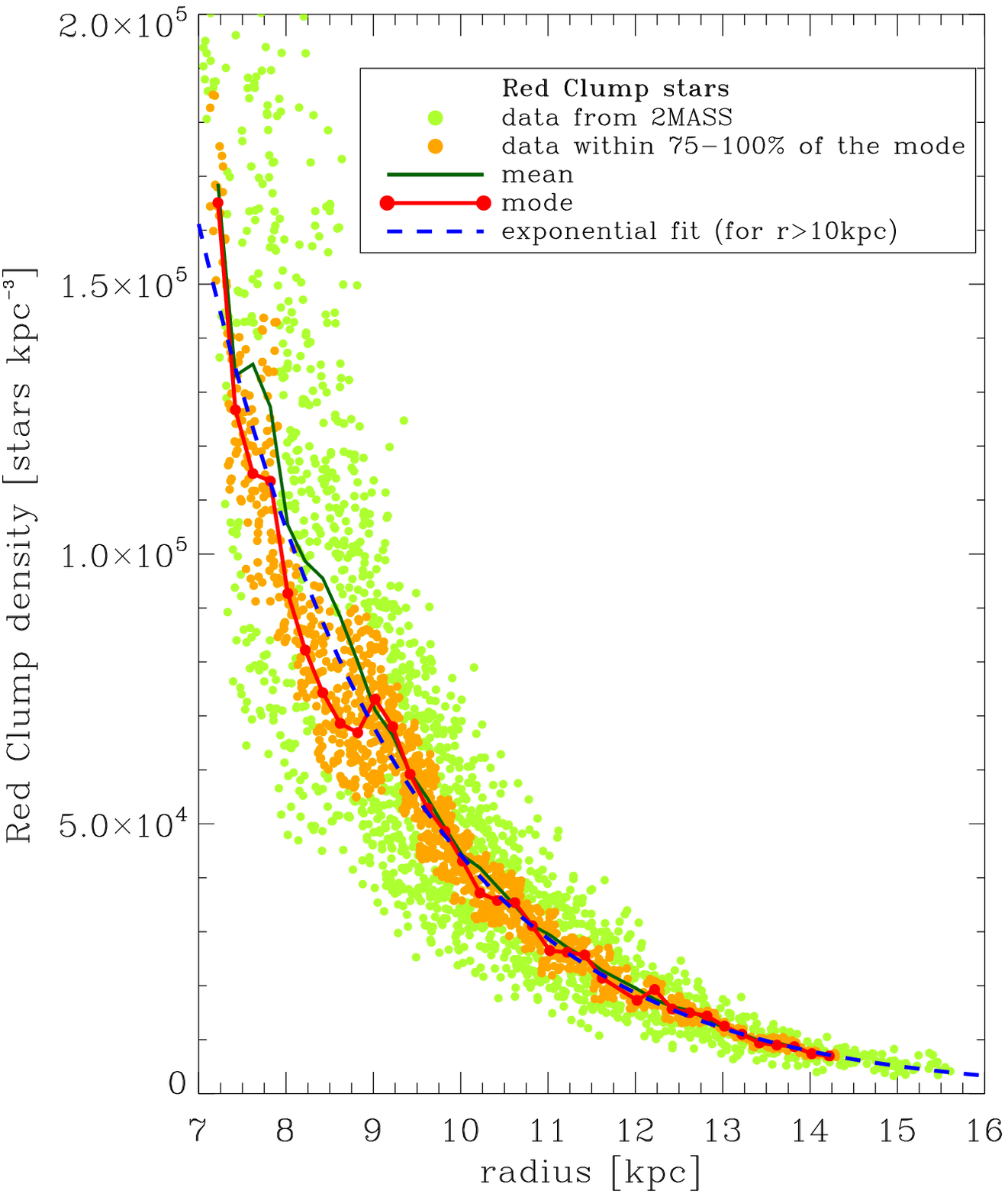}
	\caption{{\bf Left panel:} Distribution of Galactic radii of open clusters: histogram with blue and red curves overlaid - clusters older than 0.5 Gyr and 1 Gyr, respectively. {\bf Middle panel:} Distribution of Galactic radii of Cepheids older than 100 Myr. The bin size of the histograms is 0.2 kpc. {\bf Right panel:} Distribution of spatial density of red clump stars as a function of Galactic radius. The filled circles show the derived radii and densities for red clump star counts from the 2MASS data; the green and red curves represent, respectively, the mean and the mode of the distributions of density for data points within bins of 0.2 kpc of Galactic radius; the orange filled circles mark the densities with frequencies within $75\%-100\%$ of the mode of the distribution at a given bin of radius.}
	\label{fig:AKDE_clust_cef}
	\end{figure*}

The direct probing of the minimum stellar density does not seem to be a straightforward task, considering the current available observational data. A good tracer of the minimum density should present characteristics that follow the criteria: (1) heliocentric distances determined with great accuracy; (2) an observed spatial distribution on the Galactic plane that encompasses a large range of Galactic azimuth $\theta$ and radius $r$.
Another important property of the tracer of the minimum density is its age. In principle, old objects would be more suitable to trace the minimum in the stellar density, since this feature is caused by the secular evolution induced by the spiral perturbation. On the other hand, young objects would be better tracers of the minimum in the gas density; the star formation rate is directly proportional to the density of gas and, additionally, young objects, in general, do not have enough time to suffer large radial migrations, staying almost at the same radius of their birth. However, one problematic issue is that the very young objects, and those ones which are near the corotation circle (with small velocities relative to the spiral pattern), will essentially trace the spiral arms structure of the Galaxy. In this case, it might be difficult to discern a feature related to the minimum in the gas density from that one associated with the lower number of objects in an inter-arm region of the spiral structure. However, since the epicyclic frequency $\kappa$ at the solar radius corresponds to a period of the epicyclic motion of $\sim 150$ Myr (\citealt{Lepine2008}), objects older than this age are more likely to have filled the inter-arm regions of the spirals and present a more uniform distribution of Galactic radii. Given all these considerations, we have analysed the Galactic distribution of samples of open clusters, Cepheids, and red clump stars, attempting to use these objects as tracers of the minimum stellar density. 
Regarding the ages, the open clusters sample covers a large range of ages, from $\sim$ 1 Myr to 10 Gyr (see, e.g., fig. 6 from L+8); using the period-age relation from \citet{Efremov2003} for the Galactic Cepheids sample, we find an age distribution ranging from about 20 to 250 Myr, with a peak around 100 Myr; the red clump stars are typically $200-300$ Myr or even older according to \citet{Carraro2013}, or their minimum age is about $\sim 0.9-1.5$ Gyr according to \citet{Tolstoy1999}.  

For the open clusters sample, we make use of the {\it New Catalogue of Optically Visible Open Clusters and Candidates\/}, published by \citet{Dias2002}\footnote{Available at the web page www.astro.iag.usp.br/$\sim$wilton}. The present 3.3 version of the catalogue contains 2174 objects, of which 1628 have published distances and ages. 
For the analysis, we have selected subsamples of clusters older than 500 Myr and also older than 1 Gyr, which respectively totalize a number of 629 and 361 objects. The Cepheids sample was obtained from the catalogue of \citet{Berdnikov2000}, which contains data for 455 Galactic Cepheids and is available at the {\it VizieR\/} database\footnote{http://vizier.u-strasbg.fr/viz-bin/VizieR}. 
Rather than using the heliocentric distances from the catalogue, we preferred to re-compute them using the corrections from \citealt{Lepine2011a}.
In any case, the difference between the new derived distances and those ones from the catalogue is only 0.27 kpc rms. 
For our analysis, we selected the Cepheids with pulsation periods smaller than 6 days, which corresponds to the stars older than $\sim 100$ Myr and totalizes 210 objects.
The Galactic radii of the open clusters and Cepheids were calculated using the adopted $R_{0}=7.5$ kpc.

The distributions of Galactic radii of the selected samples of open clusters are shown in the left panel of Fig.~\ref{fig:AKDE_clust_cef}, in terms of histograms with bin sizes of 0.2 kpc. The top histrogram shows the distribution for objects with ages $\geq 0.5$ Gyr, and the bottom one for ages $\geq 1$ Gyr. Striking local minima in the number of counts of objects can be observed in both histograms, at radii $\sim$ 7.6 and 7.8 kpc.   
By means of Monte-Carlo simulations, we investigated the influence of the errors on the heliocentric distances of the clusters on the structure of the minimum in the counts: we constructed samples of open clusters with heliocentric distances obtained from the actual estimated distances and to which random errors, normally distributed and with a dispersion of 15\% (1$\sigma$), were added. The choice of this standard error was based on the statistics presented by \citet{Paunzen_Netopil2006} that about 80\% of the open clusters present absolute errors of the mean distances less than 20\%. We simulated a number of 100 samples, and for each one of them an adpative kernel density estimate was constructed: Gaussian kernels were used to construct the density estimator function and the optimal value for the bandwidth $h$ was determined using the least-squares cross-validation method (\citealt{Skuljan1999,Silverman1986}). This procedure is thought to be able to eliminate spurious features in the density estimation, preserving only the modes that seem to be really present. All the distributions obtained at the end of these simulations showed the presence of the minimum in the density of clusters and approximately at the same positions of the ones shown in the left panel of Fig.~\ref{fig:AKDE_clust_cef}; the solid curves that overlaid the histograms are the average density distributions after the application of the Monte-Carlo test. Since the minimum of counts observed in the distribution of old open clusters is very close to the solar circle, we argue that the completeness effects pointed out by \citet{Bonatto2006} must not affect the observed counts of objects at such radii. Indeed, we verified that a minimum of counts is also present when we select the old clusters distant from the Sun of up to 1.5 kpc. These tests led us to conclude that the minimum in the radial distribution of old open clusters is a statistically significant feature. 
As we discussed before, the ages of these objects 
rule out the hypothesis of the minimum in the counts being a mere feature of an inter-arm region of the spiral structure. An inspection of the plot of the two-dimensional distribution of these objects on the Galactic plane does not reveal any discernible structure based on the Galactic spiral arms.

The middle panel of Fig.~\ref{fig:AKDE_clust_cef} shows the distribution of Galactic radii of the Cepheids sample. It can also be observed a minimum in the counts of stars similar to the one presented by the old open clusters, in this case at an average radius of 8 kpc. Despite this good concordance, since the Cepheids are relatively young objects, it is unlikely that the minimum of counts in the Cepheids sample is the result of the secular processes studied in this paper, which drives the long-term evolution of an old population. Also unlikely is its association with an inter-arm region, since the sample is composed by objects older than 100 Myr. It could be a reflexion of the minimum in the gas density at the corotation circle (ALM), then converted into the minimum of young stars, which have not been able to fill the gas gap by their orbital scattering because some of them have been trapped on the horseshoe orbits, where the stars spend more time alternating between regions inside and outside corotation and crossing quickly this resonance radius (\citealt{Lepine2003}).


We also attempted to analyse the radial density distribution of red clump stars extracted from the 2MASS\footnote{http://www.ipac.caltech.edu/2mass/} near infrared point sources survey (\citealt{Skrutskie2006}). The method used here is identical to that developed by \citet[hereafter L02]{Lopez-Corredoira2002}, which consists in the isolation of the red clump giant stars in the color-magnitude diagrams ($K_{S}$ vs. [$J-K_{S}$]) and the inversion of its star counts to obtain the density distribution along the line of sight. Since all the details of the method are described in $\S$ 3 of L02, we do not repeat them here. We extracted the 2MASS photometric data for a total of 94 selected regions in the Galactic plane ($b=0^{\circ}$) and in the longitude range $70^{\circ}\leq l \leq 290^{\circ}$. 
The areas covered by the fields vary between 1.5 and 5 square degrees, without overlapping. The only modification to the method introduced here is that we increased the lower limit of the apparent magnitude of the red clump strip to be extracted, in most of the cases to the value $K_{S}=10.0$ (depending on the field observed), instead of the fixed value $K_{S}=8.5$ used in L02. This may, in principle, reduce the Poisson errors which mainly affect the counts of close stars, as commented by \citet{Lopez-Corredoira2004}. For each line of sight the density of stars as a function of distance was computed, and its dependence on the Galactic radius is shown in the right panel of Fig.~\ref{fig:AKDE_clust_cef}, for all lines of sight observed. In the panel, each filled circle (green and orange) shows the measured density of red clump stars at a given Galactic radius and for a given line of sight. The solid green curve and the red curve with dots and dashes represent, respectively, the mean and the mode of the distributions of density for data points within bins of 0.2 kpc of Galactic radius. The measured densities whose frequencies are within $75\%-100\%$ of the mode of the distribution at a given bin are marked as orange filled circles. Since the dispersion of points is smaller at larger radii, we attempted to fit the distribution of data points with radii $r>10$ kpc by the simple exponential law $\rho_{*}\propto e^{-r/r_{d}}$, with the scalelength $r_{d}=2.32$ kpc. The fit (blue dashed curve in the figure) performs well for the mean and mode of the distributions at least for radii greater than 10 kpc. It can be noticed that the distribution of the modes presents a systematic deviation from the exponential fit, with lower values of densities, at radii in the range $\sim 7.8$ to $\sim 8.8$ kpc. Moreover, almost all the densities within $75\%-100\%$ of the modes and between $\sim 7.8$ and 8.5 kpc have values below the exponential fit. On the other hand, the distribution of mean densities shows an opposite trend, with densities above the exponential fit for the same range of radius. This is probably due to the high density dispersion at such radii, which makes the mean of the distributions more sensitive to the high density values of the outlier points. If we were brought to believe that the distribution of the modes is more likely to give the expected dependence of the density of red clump stars as a function of Galactic radius, then we could associate the lower values of densities between $\sim$ 7.8 and 8.8 kpc described above with the predicted ring of minimum stellar density at the corotation circle. However, the relative amplitude os such decrease in density is only about 15\%, almost half of the expected value. Of course, the high scattering of the data points in the figure shows us that it does not seem to be cautious to draw any conclusion about relatively small scale features in the underlying stellar density of the disk. We could even point out a possible source of systematic error: the dependence of the estimated distances on the metallicity of the stars, in conjuncion with the observed step decrease in the abundance pattern near the corotation radius (see below), that could be leading to the apparent break in the exponential profile of the distribution of the modes of density at radii between 8.5 and 9 kpc, visible in the right panel of Fig.~\ref{fig:AKDE_clust_cef}. In any case, further work in this field is needed to establish the best way of using the red clump stars as tracers of the minimum stellar density at corotation.

The old open clusters sample seems to give the strongest evidence for the existence of a small-density ring in the stellar component at the corotation radius of the Galactic disk. The radius where such minimum of counts occur ($r\sim 7.6-7.8$ kpc) is slightly smaller than the average radius of 8.3 kpc of the gap in the H\,{\sevensize I} density found by ALM. 
Although this is a minor difference, it could be related to the fact that the minimum of gas density can be slightly shifted with respect to the corotation radius as revealed by hydrodynamic simulations (see fig. 4 of \citealt{Lepine2001}). Accordingly, the gap in the density of
very young objects should be related to the minimum of gas density rather than to the exact corotation radius, simply because stars cannot born in the absence of gas. In this way we may also understand some other minor discrepancies. 
An abrupt step decrease of 0.3 dex in the radial distribution of [Fe/H] abundance of open clusters was found in the Galactic disk by \citet*{Twarog1997} and was confirmed by L+8 using more recent data. Adopting $R_{0}=7.5$ kpc, the radius of the sharp discontinuity in the metallicity distribution is at about 8.5 kpc (see, e.g., fig. 5 from L+8 or fig. 1 from \citealt{Scarano_Lepine2013}).  
As explained by L+8, the metallicity discontinuity is itself a consequence of the ring-shaped gap in the density of gas, which isolates the inner and outer sides of the gap one from the other and leads to an independent evolution of the metallicity on the two sides. The connection between the radii of metallicity discontinuities and corotation radii was further investigated by \citet{Scarano_Lepine2013} using a large sample of spiral galaxies. They found a clear correlation between these radii, but with a tendency of the galactic radii at which breaks or changes of slope of the metallicity gradients occur to be slightly larger (10\%) than the corotation radii. One possible cause of this deviation could be the shift of the gas minimum that we already mentionned. 
Coming back to our Galaxy, we have to keep in mind that the indirect measurements based on metallicities give values for the corotation radius that are within the expected uncertainty of 0.6 kpc (for $R_{0}=7.5$ kpc), as derived by \citet{Dias_Lepine2005}.



\section[]{Additional remarks}
\label{add_remarks}

\begin{enumerate}

\item {\it Changes in the initital conditions of the stellar disk and the axisymmetric potential\/}. We have run simulations with different values for the disk scalelength $r_{d}$, such as 2.0, 3.0 and 3.5 kpc, and with the same setup of parameters of the Galactic potential used in the simulation with the spiral model Sp1. Essentially, the same configurations of the final distribution of surface density were obtained compared to that from the model Sp1. 
We also have checked the condition of the initial axisymmetric potential being derived from a rotation curve with the minimum at $\sim 9$ kpc being already present at the beginning of the simulation. This condition relies on the hypothesis of the minimum in the rotation curve being associated with another property of the Galactic potential, instead of the minimum of stellar density at corotation. However, a configuration of the relative density variations similar to the one shown in Fig.~\ref{fig:distr_dL_amprel}d was obtained, which shows that the formation of the minimum density depends more strongly on the properties of the spiral perturbation.

\item {\it Rings in the outer disk of the Galaxy\/}? The density variations resultant from the simulations using the different models of spiral potential, which are shown in Figs.~\ref{fig:distr_dL_amprel}d and~\ref{fig:KDE_amprel_distrib}, show a similar pattern in the outer regions of the disk: local minima density at the 2:1 and 4:1 outer Lindblad resonances, and a local maximum density between these resonances and beyond the 2:1 OLR. In some cases, the amplitudes of these minima and maxima density reach a similar value to that of the minimum at corotation. However, they must be formed by a different process. We have seen that during the stages of linear growth of the perturbation, stars located near the outer resonances gain angular momentum from the wave, which puts them in orbits with larger radii. Part of the amount of orbital energy gained in this process is converted into non-circular motion. This can be checked from Fig.~\ref{fig:sigVr_simulat}, where peaks in the radial velocity dispersion are observed at the 2:1 and 4:1 OLRs. As the amount of $L$ exchanged in the outer disk, at the steady state of the wave, is smaller compared to that in the inner disk, the low radial scattering of the orbits is not efficient in blurring the features in the stellar density that had been created at the earlier stages, which seems to occur in the inner disk. Thus, the quasi-circular behaviour of the orbits in the outer disk helps to maintain the ringed pattern of the density that we observe from the simulations. We emphasize that such ringed pattern strongly depends on the properties of the spiral structure in the outer disk. Therefore, this result must be treated with some skepticism, since our models make use of an extrapolation for large radii of the local observable properties of the spiral arms. That is not the case of the corotation region, since due to its proximity to the solar circle, the more well defined parameters of the spiral arms lead to a minimum density at corotation whose existence is more reliable than the others in the outer disk. Nevertheless, the existence of rings in the outer Galaxy has been considered by \citet{Binney_Dehnen1997} as a possibility to explain the apparent rise of the rotation curve at radii $r \ga 1.25 R_{0}$. The errors on the distances estimates of tracers concentrated into such rings could lead to the result $V_{c}\propto r$, mimicking a disk with rigid body rotation. The authors calculated a high probability of most of the tracers lying in a ring of radius $\approx 1.6 R_{0}$, which in our adopted scale results in $r=12$ kpc. This radius is close to the region where an overdensity between the 4:1 and 2:1 OLRs appears in our simulations, with an average radius between 11.5 and 13 kpc. It is also close to the radii where breaks in the metallicity gradient of open clusters (\citealt{Yong2012}) and Cepheids (\citealt{Andrievsky2004}; L+8) occur in the Galactic disk. In the case of external galaxies, ALM stated that ring-shaped gaps are present in the H\,{\sevensize I} density profile of a number of objects studied in the literature. As an example, \citet{Schommer_Sullivan1976} argued for the relation between the resonance regions and the ring-like structures observed in the spiral galaxy NGC 4736. The authors showed that the position of the 2:1 ILR is coincident with a ring-like zone of H\,{\sevensize II} regions, the corotation is associated with a gap in the optical brightness and in H\,{\sevensize I}, and also the 2:1 OLR coincides with the position of a faint outer ring of stars and H\,{\sevensize I}, and also possibly a ring of H\,{\sevensize II} regions. Later on, \citet{Gu1996} conducted gas cloud-particle simulations and showed that both inner and outer rings in NGC 4736 are stable structures located respectively at the ILR and OLR.

\item {\it The lifetime of the spiral pattern\/}. The question of whether the spiral arms are short- or long-lived structures is still a matter of debate (e.g. \citealt{Sellwood2011,Siebert2012}). In the case of the Milky Way, some recent observational constraints that tend to favour a long-lived picture for the spirals have been put in evidence. For instance, L+8 estimated the time required to build up the step in the metallicity distribution of open clusters in the Galactic disk as being of the order of 3 Gyr. This value is a measure of the minimal lifetime of the present grand design spiral pattern, which means that the corotation resonance stayed at the same position during this period of time. This result is also in agreement with the work from \citet{Maciel2003}, which show that Galactic planetary nebulae younger than 4 Gyr have a radial metallicity distribution with a flat behaviour or even a positive slope beyond corotation. This result is compatible with the idea that the star formation rate in the disk is also proportional to the relative velocity of the gas with respect to the spiral arms, SFR $\propto |\Omega-\Omega_{p}|$ (e.g. \citealt{Mishurov2002}, among others). Our result in the present paper that the elapsed time of 3 Gyr, required to build up a minimum in the stellar density at corotation of $\sim 30\%$ to 40\% of the underlying density, comes as another indirect evidence for such lower limit to the age of the spiral structure of the Galaxy.

\item {\it A ring of minimum stellar density near the solar orbit radius\/}. Due to the proximity of the corotation resonance, the solar orbit would be placed very close to the minimum of the disk density. This would result in strong implications to the measured density and kinematics of the solar neighbourhood, as well as to the Galactic disk as a whole. To investigate such implications, we constructed a model of an exponential disk with scalelength $r_{d}=2.5$ kpc and a surface density to which a minimum followed by a maximum were added, with the same radial profile shown in Fig.~\ref{fig:surfdens_rotcurv} by the red curve and expressed by Eq.~\ref{eq:Sigma_function}. We chose the same parameters of the function $f_{\mathrm{mrc}}$ as given in $\S$~\ref{potaxis}: $R_{\mathrm{mrc}}=8.9$ kpc, $\sigma_{\mathrm{mrc}}=0.8$ kpc, and an amplitude $A_{\mathrm{mrc}}$ that multiplied by the constant $\mathcal{C}$ gives a factor of 0.45. This radial profile for the surface density puts the corotation at $R_{cr}=8.1$ kpc and with a relative amplitude of the minimum density of 34\% of the underlying density. At the solar Galactocentric distance, $R_{0}=7.5$ kpc, the relative density is 21.3\% lower than the underlying density. This model implies that the disk would be in average more than 20\% denser at radii before and beyond $R_{0}$ than a disk without the minimum density at corotation, considering the same normalization for the local surface density $\Sigma_{0}$. This difference in density is represented in Fig.~\ref{fig:surfdens_rotcurv} by the vertical shift between the dashed and dotted red straight lines. Setting $\Sigma_{0}=50$ M$_{\odot}$ pc$^{-2}$ at the solar radius, the total mass of the disk presenting the density variations at the corotation circle would be about 20 \% larger than presently estimated, the equivalent of a disk with $\Sigma_{0}=60$ M$_{\odot}$ pc$^{-2}$.
 This result puts us in the following situation: 
if the solar orbit radius is indeed inside a ring of minimum density and this fact is not taken into account in the determination of the disk mass, we are neglecting a considerable fraction of mass in the disk component of the Galaxy. Considering now the Galactic disk model of SHO, and with their parameters for the ring of density beyond $R_{0}$ (which is similar to our profile in Eq.~\ref{eq:Sigma_function}), the above quoted difference in mass is as large as 39\% of the mass of the disk without ring (in this calculus we consider the same $\Sigma_{0}$ for both disks with and without the ring of density, modeled by the authors as $\Sigma_{0}=87.5$ M$_{\odot}$ pc$^{-2}$). These results also have strong implications to the contribution of the disk to the rotation curve of the Galaxy. As an example, \citet{Lepine_Amaral1999} found a difference of about 30\% between the local surface density derived from star counts plus a gaseous disk and the one needed for the fit of the rotation curve. \citet{Sackett1997}, based on new observational constraints on the structure of the Galactic disk, verified that the `maximal disk hypothesis' commonly applied to external galaxies, also gives a maximal disk when applied to the Milky Way. According to this definition, to be maximal, the disk must provide $85\%\pm 10\%$ of the total rotation support of the galaxy at the radius $r=2.2\,r_{d}$. Using our disk model ($\Sigma_{0}=50$ M$_{\odot}$ pc$^{-2}$; $r_{d}=2.5$ kpc; $R_{0}=7.5$ kpc),
we estimate that the disk with and without the minimum density at corotation is responsible for 84\% and 74\% of the total circular velocity at $r=5.5$ kpc, respectively, which puts them in the expected range to be considered as maximal disks. If we considered a higher rotation curve, with $V_{c}(5.5\,\mathrm{kpc})\approx 230$ km s$^{-1}$ for instance, the disk with the minimum density at corotation would provide 80\% of the total circular velocity at $r=5.5$ kpc, still being maximal. We argue here that the assumption that the solar orbit lies close to a ring of minimum density implies in a correction to larger values for the total mass of the Galactic disk, and consequently, a greater contribution of the disk component to the inner rotation curve of the Galaxy. This result also implies in a dark halo less important to the dynamics of the inner Galaxy, or would even rule out models of the Galaxy that evoke a dark matter disk with the same structural parameters of the observed luminous disk. We finish pointing out that the minimum density at corotation does not seem to be, in principle, incompatible with Galactic models based on star counts in infrared bands. For instance, Polido et al. (2013, submitted) performed star counts model of the Galaxy using the 2MASS data and verified that a disk model with a minimum density close to $R_{0}$ generates a little better fit to the counts in the $K_{\mathrm{\scriptstyle S}}$ band in the Galactic plane than a disk model without the minimum density. In agreement with the results shown above, their disk model with the correction in the density leads to an increase in the theoretical counts, in all range of Galactic longitudes, when compared to that from the disk without the correction, which means that a more massive disk is obtained when the minimum density is taken into account.

\end{enumerate}



\section{Conclusions}
\label{outro}

We have presented arguments for the existence of a minimum in the disk stellar density at the corotation radius of the spiral structure of the Galaxy. The formation of such minimum is expected to be driven by the secular evolution of the Galactic disk induced by the exchanges of energy and angular momentum between the stars and the spiral density wave. The resulting redistribution of the disk stellar density is achieved by the secular decrease and increase in the mean orbital radius for stars inside and outside corotation, respectively. The mechanism that governs this process has been elucidated by Z96, who showed that the flow of the disk matter in opposite directions from corotation is the result of the torque applied by the spiral potential on the stellar orbits that contribute to the spiral density, and the sign of the torque is reversed from one to the other side of corotation.

Despite the fact that our simulations of stellar orbits under the Galactic gravitational field give support to the formation of the minimum stellar density at corotation, we have observed a degeneracy in the parameter space of the spiral potential that leads to similar results for the characteristics of the minimum density. However, we have verified that, given the range for the estimated amplitude of the spiral perturbation of the Galaxy, a minimum in the stellar density with a relative amplitude of $\sim 30\%$ to 40\% of the underlying density can be formed at corotation. We have also investigated the influence of multiple spiral patterns, as well as the interactions between spirals and a central bar, to verify the efficiency of the disk heating at the solar radius. Although our tests favour the scenario of two steady spiral patterns with different angular speeds for both the formation of the minimum density and the observed age-velocity dispersion of the solar neighbourhood, we rise the hypothesis that other mechanisms of stellar heating could have been operating in conjunction with the heating from a single spiral pattern, such as, for instance, the infall of small satellite galaxies in the Milky Way disk. 

A minimum in the surface density is compatible with the observed local dip in the Galactic rotation curve between 8.5 and 9.0 kpc (for $R_{0}=7.5$ kpc). Such dip in the rotation curve appears as a strong evidence for the presence of the minimum density at corotation, not only in the stellar but also in the gas disk component. The minimum in the gas density and in the young stellar populations had already been probed by ALM. In the present paper, we deal with the minimum density in the old stellar population, which can be reflected in the dip of the rotation curve. These observational features put the corotation resonance very close to the solar orbit radius, an argument that has been stated by several authors during the past decades. 
We also report the presence of a striking minimum in the number of counts of a sample of open clusters older than 1 Gyr, placed at the radius $r \sim 7.7$ kpc. We associate this minimum with the predicted ring of small stellar density at corotation. These Galactic features are indicators of a spiral pattern with a long-lived structure, with its corotation radius being kept constant over a few billion years. 


A Galactic ring of minimum stellar density at corotation is also reproduced when we consider the spiral arms as being caused by the crowding of stellar orbits in the disk. In fact, a particle simulation using the spiral potential with Gaussian azimuthal profile from JLBB generated a minimum stellar density quite similar to the ones using the commonly employed cosine profile. This represents one step forward in our understanding of the nature of the spiral perturbation; the description of the perturbations as narrow grooves or channels in the gravitational potential with an approximately Gaussian profile has been giving a self-consistent picture for the spiral structure.

The hypothesis of the solar orbit lying inside a ring of minimum density has a strong implication to the estimated total mass of the Galacitc disk. Considering the current estimated values for the local surface mass density and the disk scalelength, we showed that a correction for a $\sim 20\%$ more massive disk would be applied to the current estimates of the disk mass of the Galaxy. This would, in turn, lead to a greater contribution of the disk component to the inner rotation curve of the Galaxy, and consequently, a less important dark halo component. 
Finally, we comment that it has been tempting to interpret the minimum in the rotation curve just beyond $R_{0}$ as being the result of a transition between a decreasing contribution from the disk and an increasing contribution from the dark halo or even a dark disk. However, it is known that such a transition would be difficult to explain in terms of the smooth distributions of the disk an dark halo masses. Contrary to this interpretation, we have shown that such dip in the rotation curve can naturally be explained by local variations in the disk surface density, such as the minimum density at corotation. 
We expect that the existence of a ring of minimum stellar density just beyond the solar orbit radius can be further investigated with the forthcoming observations such as those from GAIA, the observations of red clump stars in the disk by the LAMOST-LEGUE survey, the APOGEE survey, among others.


\section*{Acknowledgments}

We acknowledge the anonymous referee whose comments and suggestions have improved significantly the present paper. This research has made use of the VizieR catalogue access tool, CDS, Strasbourg, France. This publication makes use of data products from the Two Micron All Sky Survey, which is a joint project of the University of Massachusetts and the Infrared Processing and Analysis Center/California Institute of Technology, funded by the National Aeronautics and Space Administration and the National Science Foundation. This work was partially supported by the Brazilian research agencies CAPES-PROEX and CNPq (140143/2012-2). 

\bibliographystyle{aa}
\bibliography{refs} 

\bsp

\label{lastpage}

\end{document}